%% file: main.tex
\documentclass[
 amsmath,amssymb,
 aps,
 superscriptaddress,
 preprint 
]{revtex4-2}

\usepackage[pdftex]{graphicx}
\usepackage{dcolumn}
\usepackage{caption}
\usepackage{tabularx}
\usepackage{accents}
\usepackage{bm}
\usepackage{braket}
\usepackage{mathrsfs}
\usepackage{booktabs}
\usepackage{array}
\usepackage{hyperref}
\usepackage{color}

\newlength\myindention
\DeclareCaptionFormat{myformat}{\hspace*{\myindention}#1#2#3}
\begin{document}

\title{Quantum Monte Carlo Calculations of neutron-$\alpha$ Scattering via an Integral Relation}
\makeatletter
\def\frontmatter@affiliationfont{\itshape\small}
\makeatother

\author{Abraham R. Flores}
\affiliation{Department of Physics, Washington University in St. Louis, St. Louis, MO 63130, USA}
\author{Kenneth M. Nollett}
\affiliation{San Diego State University, San Diego, CA 92182, USA}
\author{Maria Piarulli}
\affiliation{Department of Physics, Washington University in St. Louis, St. Louis, MO 63130, USA}
\affiliation{McDonnell Center for the Space Sciences at Washington University in St. Louis, MO 63130, USA}
\date{\today}

\begin{abstract}
Nuclear physics seeks to describe both bound and unbound states within a unified predictive framework. While coordinate-space Quantum Monte Carlo (QMC) methods have successfully computed bound states for systems with $A \leq 12$, their application to unbound states remains limited. In this work, we extend the QMC approach to enable a broader range of unbound-state calculations. Our method infers long-range amplitudes in the wave function from integrals over the short-range interaction region. By evaluating these integrals using Green's Function Monte Carlo wave functions with the Argonne $v_{18}$ potential, we accurately reproduce existing results for neutron-alpha scattering. This approach provides a systematic pathway for studying more complex nuclear systems, including coupled-channel scattering and the effects of three-nucleon forces. It serves as a powerful tool for advancing {\it ab initio} calculations in nuclear reactions, paving the way for a unified framework that consistently describes both bound and scattering states within a single theoretical approach.
\end{abstract}

\maketitle

\input{Sections/Introduction}
\input{Sections/QMC}
\input{Sections/Methods}
\input{Sections/Results}
\input{Sections/Conclusion}

\begin{acknowledgments}
We would like to thank R.\ B.\ Wiringa for being instrumental in the process of generating and applying the variational wave functions, S.\ Pastore for a smooth transition to St. Louis, A.\ Lovato for useful discussions of the GFMC mixed estimates, M.\ Viviani for a detailed explanation of the regularization functions, and M.\ Paris and G.\ Hale for explanations of the $R$-matrix phase shifts. This work was supported by the US Department of Energy, Office of Nuclear Physics, 2021 Early Career Award number DE-SC0022002 (A.R.F and M.P.), Award DE-SC0019257 (K.M.N.), FRIB Theory Alliance award DE-SC0013617 (M.P.), and the NUCLEI SciDAC program (A.R.F and M.P.). The many-body calculations were performed on the parallel computers of the Laboratory Computing Resource Center, Argonne National Laboratory, and the computers of the Argonne Leadership Computing Facility via the INCITE grant ``Ab initio nuclear structure and nuclear reactions''. This research used resources of the National Energy Research Scientific Computing Center, a DOE Office of Science User Facility supported by the Office of Science of the U.S. Department of Energy under Contract No. DE-AC02-05CH11231 using NERSC award NERSC DDR-ERCAP0026818.
\end{acknowledgments}

\bibliography{QMCna}

\end{document}

%% file: Sections/Introduction.tex
\section{\label{section:intro}Introduction}

A fundamental goal of nuclear physics is to achieve a microscopic understanding of nuclear phenomena through the interactions of nucleons. Over the past three decades, significant progress has been made toward this goal with the development of \textit{ab initio} many-body methods~\cite{Hergert:2020bxy, Ekstrom:2022yea, Machleidt:2023jws, Navratil:2022lvq}, which provide approximate solutions to the nuclear Schrödinger equation, and the systematic formulation of nuclear forces using Effective Field Theory (EFT)~\cite{Weinberg:1990rz,Weinberg:1991um,Weinberg:1992yk,Ordonez:1992xp,Rho:1990cf,Epelbaum:2008ga,Hammer:2019poc,Machleidt:2024bwl}. These advances, coupled with the integration of uncertainty quantification techniques~\cite{Furnstahl:2014xsa,Phillips:2020dmw}, have propelled low-energy nuclear theory into an era of precision calculations, enabling rigorous computational simulations that complement experimental investigations and enhance the reliability of nuclear physics inputs in astrophysics, particle physics, and other domains.

Although \textit{ab initio} nuclear structure calculations now extend to medium-mass and even heavy nuclei~\cite{Epelbaum2011Hoyle, Novario2021CC,hu2022Pb208, elhatisari2024wavefunction,Navratil:2007we,Roth:2011ar, Roth:2011vt, Hagen:2012sh,Hagen:2012fb, Barrett:2013nh, Hergert:2012nb, Hagen:2013nca, Binder:2013xaa, Hagen:2016uwj, Simonis:2015vja,Simonis:2017dny, Morris:2017vxi,Soma:2019bso,Hoppe:2019uyw,Huther:2019ont,Lonardoni:2017egu,Binder:2013xaa,Hagen:2015yea,Hu:2021trw,Novario:2021low,Sammarruca:2022rcl}, their application to nuclear reactions and scattering problems remains significantly more challenging. This is due to the need for an accurate treatment of continuum states—a requirement that complicates numerical implementation and limits existing \textit{ab initio} methods to only a handful of systems~\cite{Zhang:2020rhz,Yang:2025chiral, CAPUZZI2000223,Escher:2002ud,Dickhoff:2016ikd,Dickhoff:2018wdd,Dickhoff:2004xx, Tostevin:2014usa,Aumann:2017jvz,Capel:2008mw,Hebborn:2018uky,Summers:2006dm,summers2014erratum,Moro:2012kw,deDiego:2013waa,Gomez-Ramos:2015fxa,Ogata:2003ah,Baye:2005ib, Fossez:2015cma,Fossez:2015qxa, kievsky2008high,marcucci2009n,barletta2009integral,kievsky2010variational,viviani2020n+, lazauskas2004testing,lazauskas2005low,lazauskas2009elastic,lazauskas2012application, Lazauskas2018Sol,lazauskas2020description,deltuva2014calculation,deltuva2015deuteron,deltuva2015four,deltuva2015proton,deltuva2017four,alt1978coulomb,alt1980scattering,deltuva2005momentum,deltuva2005calculation,Navratil:2010jn}. Unlike bound-state problems, nuclear scattering calculations must account for asymptotic boundary conditions, the presence of open decay channels, and the need for precision in describing reaction dynamics. These factors impose substantial computational demands, making large-scale \textit{ab initio} reaction studies difficult~\cite{Johnson2020White,LRP2023}.

However, understanding nuclear reactions is essential to address fundamental scientific questions such as the origin of elements and the evolution of the universe. Low-energy fusion reactions drive stellar nucleosynthesis, governing energy production in stars, and shaping the formation of chemical elements. Furthermore, nuclear reactions provide a critical tool for probing exotic nuclei—rare isotopes at the limits of stability, many of which have unbound low-lying states that cannot be rigorously analyzed without an accurate treatment of continuum effects.

QMC methods have been successfully applied to study properties of bound state systems up to $A \leq 12$~\cite{pudliner1997quantum,Pieper2009QMC,Pieper2002Quant,piarulli2015minimally,piarulli2016local,piarulli2018light,piarulli2020benchmark,Pastore2009Elec,Pastore2018Quant,King2020Weak,King2020Chiral,King:2023ab,King:2022Partial,Chambers-Wall:2024fha,King:2024jiq,Andreoli:2024ovl,Lovato:2023raf,Lovato:2023khk} since they are limited by the exponential growth of spin-isospin states. However, leveraging recent advances in computational resources, the QMC approach presents a promising avenue to extend up to $A = 13$ systems \cite{Flores:inprog:A13}. On the other hand, the application of QMC to nuclear scattering problems has been more limited due to the challenges in treating continuum states and enforcing asymptotic boundary conditions. Unlike bound-state calculations, where wavefunctions vanish at large distances, scattering problems require precise handling of wavefunction tails and phase shifts, making their numerical implementation significantly more complex.

The first applications of QMC methods to scattering problems included studies of proton and neutron scattering from tritium and alpha particles using the Variational Monte Carlo (VMC) method~\cite{carlson1984variational, carlson1987microscopic}. Later, the same method was extended to the Green's Function Monte Carlo (GFMC) to calculate the neutron-alpha scattering using phenomenological potentials~\cite{nollett2007quantum}. Subsequently, neutron-alpha scattering calculations were further developed using local chiral Darmstadt interactions~\cite{lynn2016Chiral}. More recently, the methods used in this work represent an extension to GFMC of Ref.~\cite{flores2023Var}, which used VMC to study neutron scattering from tritium. Details of this extension are provided in the following sections.

The QMC approach to scattering employs a spherical particle-in-a-box framework, which enforces the scattering boundary condition at the surface of the box. Specifying this boundary condition determines the relative energy between the daughters, but the total energy must still be computed using conventional QMC techniques. Then the phase shift can be calculated directly from the boundary condition and the energy for single-channel scattering; we call this the ``direct method''. The direct method, in principle, states the phase shift and calculates the energy; this is backward to the usual scattering calculation, which states the energy and computes the phase shift. 

The difficulty that arises with the direct method is that in QMC, the tails of the wavefunction must accurately capture the collective behavior of the clusterized system. In GFMC, achieving this requires an iterative procedure to ensure the tails of the wavefunction converge to the correct asymptotic form. Moreover, the direct method is limited to single-channel scattering and cannot be extended to coupled-channel problems. To address coupled-channel scenarios, an alternative approach is necessary.

The method employed in this work is closely related to the partial-wave Lippmann-Schwinger equation \cite{lippmann1950variational} and was initially applied to scattering problems via the Kohn-Variational principle \cite{harris1967expansion}. It also shares similarities with the source term calculations of Refs.~\cite{pinkston1965Form,kawai1967Amethod} and asymptotic normalization coefficient calculations of Refs.~\cite{mukhamedzhanov1990Micro,timofeyuk1998One,timofeyuk2010Overlap,nollett2011asymptotic,nollett2012ab}. This work closely follows the notation and formulation laid out by the scattering calculations of Refs.~\cite{barletta2009integral,kievsky2010variational, Romero2011General,viviani2020n+,flores2023Var}. In this paper, we refer to this procedure as the ``integral method, '' which infers long-range amplitudes of the wave function by integrals over the region where all the particles interact. From these amplitudes, all other scattering information follows. Using this integral framework, we minimize the influence of wavefunction tails on the calculated observables, enhancing the reliability of the results.

In this study, we validate the integral method through neutron-alpha ($n\text{-}\alpha$) scattering, leveraging existing Argonne 
$v_{18}$ (AV18) results: a previous GFMC calculation~\cite{nollett2007quantum}, and a solution to the $A=5$ Faddeev-Yakubovsky equations~\cite{Lazauskas2018Sol}. The $n\text{-}\alpha$ system is an ideal test case because, as an $A=5$ system, it is computationally feasible and involves a single-channel process up to the $d+{}^3\text{H}$ threshold, which lies at a relatively high energy. Additionally, for systems with $A \leq 5$, exact solutions to few-nucleon continuum states have been extensively studied and benchmarked using the AV18 Hamiltonian~\cite{viviani2011benchmark, viviani2017benchmark}, providing a reliable basis for comparison.
 Among these, the hyperspherical harmonic (HH) methods ~\cite{kievsky2008high,marcucci2009n,barletta2009integral,kievsky2010variational,viviani2020n+} and the Alt-Grassberger-Sandhas equations~\cite{deltuva2014calculation,deltuva2015deuteron,deltuva2015four,deltuva2015proton,deltuva2017four,alt1978coulomb,alt1980scattering,deltuva2005momentum,deltuva2005calculation} have been widely employed for $A \leq 4$ systems. The Faddeev-Yakubovsky equations have been successfully used to describe scattering up to $A=5$ \cite{lazauskas2004testing,lazauskas2005low,lazauskas2009elastic,lazauskas2012application, Lazauskas2018Sol,lazauskas2020description}. Additionally, the No-Core Shell Model combined with the Resonating Group Method (NCSM/RGM) has been effectively used to study nucleon-nucleus scattering~\cite{Navratil:2010jn}. In particular, NCSM/RGM calculations of $n\text{-}\alpha$~\cite{Quaglioni:2008sm, Quaglioni:2009mn, Kravvaris2024xcp} have  applied chiral interactions and low-momentum interactions ($V_{\text{low}k}$) derived from AV18; because those results do not use the bare AV18 interaction, we will not compare with these calculations here.

The paper is organized as follows.  In Sec.\ref{section:QMC}, we describe the VMC and GFMC wavefunctions, with a particular focus on the implementation of scattering boundary conditions. Sec.\ref{section:methods} outlines the calculation of scattering observables and the underlying assumptions of the intergal method approach. Sec.~\ref{section:reg} discusses the regularization needed for the irregular Coulomb function and its impact on the phase shifts.  Our results are presented and analyzed in Sec.\ref{section:results}. Finally, in Sec.~\ref{section:conclusion}, we summarize our findings and provide an outlook on future applications of the integral method in QMC.

%% file: Sections/QMC.tex
\section{\label{section:QMC}Quantum Monte Carlo}
QMC methods are powerful computational techniques designed to solve the non-relativistic many-body Schrödinger equation:
\begin{equation}
  \label{eq:schroedinger}
  H\ket{\Psi(\mathbf{R};J^\pi,T,T_z)} = E\ket{\Psi(\mathbf{R};J^\pi,T,T_z)},
\end{equation}
where \(E\) is the energy eigenvalue, \(\mathbf{R} = \{\mathbf{r}_1, \mathbf{r}_2, \ldots, \mathbf{r}_A\}\) represents the particle coordinates, \(J\) is the total angular momentum, \(\pi\) is the parity, \(T\) is the total isospin, and \(T_z\) is the isospin projection. The Hamiltonian \(H\), used in these calculations, can include contributions up to three-body forces:
\begin{equation}
  \label{eq:hamiltonian}
  H = -\sum_{i=1}^A \frac{\hbar^2}{2m}\nabla_i^2
  + \sum_{i< j}^{A}v_{ij}
  +\sum_{i< j< k}^{A}V_{ijk}.
\end{equation}
Here, the first term represents the kinetic energy of the particles, \(\nabla_i^2\) is the Laplacian operator for particle \(i\), \(v_{ij}\) is the two-body potential describing interactions between particles \(i\) and \(j\), and \(V_{ijk}\) is the three-body potential that accounts for interactions involving particles \(i\), \(j\), and \(k\). 

QMC methods have been particularly effective in calculating bound state properties for nuclei up to \(A \leq 12\)~\cite{Carlson2015QMC,pudliner1997quantum,Pieper2009QMC,Pieper2002Quant,piarulli2015minimally,piarulli2016local,piarulli2018light,piarulli2020benchmark,Pastore2009Elec,Pastore2018Quant,King2020Weak,King2020Chiral,King:2023ab,King:2022Partial,Chambers-Wall:2024fha,King:2024jiq,Andreoli:2024ovl,Lovato:2023raf,Lovato:2023khk,King:2024zbv,Lovato:2019fiw}, including energies, beta decay, muon capture, charge radii, momentum distributions, spectroscopic factors, and others. However, extending these techniques to describe unbound systems introduces unique challenges, such as adequately handling continuum states, offering opportunities to further refine and expand these computational approaches.
\input{Sections/VMC}
\input{Sections/GFMC}

%% file: Sections/VMC.tex
\subsection{\label{subsection:VMC}Variational Monte Carlo}
The VMC method is the first step in a typical QMC calculation, involving the construction of a variational wave function $|\Psi_V\rangle$. This VMC wave function is constructed from two- and three-body operator correlations acting on a Jastrow wave function, $\Ket{\Psi_J}$, which includes scalar correlations and a spin-isospin Slater determinant, $\Phi$.  The form of the variational ansatz is given by~\cite{wiringa2009}
\begin{equation}
    \label{varwf}
    \Ket{\Psi_V} = \left[\mathcal{S}\prod_{i<j}\left(1+U_{ij}
      +\sum_{k\neq i,j}U_{ijk}\right)\right]\Ket{\Psi_J},
\end{equation}
where the sums and products run over nucleon labels. $\mathcal{S}$ is the symmetrization operator with two- and three-body correlation operators, $U_{ij}$ and $U_{ijk}$, respectively. In the Jastrow wave function $|\Psi_J\rangle$, each nucleon is assigned to either the $s$-shell or $p$-shell depending on the permutation of particle labels. The central pair correlations are chosen accordingly—$f_{ss}(r_{ij})$ is used when both particles $i$ and $j$ are in the $s$-shell core, while $f_{sp}(r_{ij})$ is applied when one nucleon is in the $s$-shell and the other in the $p$-shell, similarly $f_{sss}(r_{ij})$ is applied when a triplet of nucleons  are all in the $s$-shell. For a nucleus with $A$ nucleons, including a fully occupied four-particle $s$-shell and only one $p$-shell nucleon, the Jastrow function takes the form:
\begin{eqnarray}
    \ket{\Psi_J} &=& \mathcal{A}\left\{ \prod_{i<j<k\leqslant 4}f_{ijk}^{sss}\prod_{t<u\leqslant 4}f^{ss}(r_{tu})\right.
    \label{jastrow}\prod_{i\leqslant 4}\hspace{1mm}\prod_{5\leqslant j \leqslant A}f^{sp}(r_{ij})\\
  &&\times\left.\sum_{LS[n]}\beta_{LS[n]}\left|\Phi_A(LS[n]JMTT_z)_P\right\rangle \right\}.\nonumber
\end{eqnarray}
The spin-isospin vector, $\Phi_A$, is given a weight $\beta_{LS[n]}$ and is coupled to specified quantum numbers of total spin $S$, orbital angular momentum $L$, and net angular momentum $J$ (with projection $M$), as well as total isospin $T$ (with projection $T_z$), as indicated by square brackets. In general, the full specification also requires a definite permutation symmetry among $p$-shell orbitals in the form of a Young diagram label $[n]$. The spin-isospin vector is defined as,
\begin{eqnarray}
    \label{phiA}
    &&\left|\Phi_A\left(LS[n]JMTT_z\right)_P\right\rangle\nonumber\\
    &&=\left|\Phi_\alpha(J=0,M_J=0,T=0,T_z=0)_{1234}\prod_{5\leqslant i\leqslant A}\phi_p^{LS[n]}(r_{\alpha i})\right.\nonumber\\
    &&\times\left[\left[\prod_{5\leqslant j\leqslant A}
        Y_{lm_l}\left(\mathbf{\hat{r}}_{\alpha j}\right)\right]_{LM_L}
      \!\otimes\left[\prod_{5\leqslant k \leqslant A}\chi_k \left(\frac{1}{2}m_i\right) \right]_{SM_S}\right]_{JM}\nonumber\\
    &&\left.\times\left[\prod_{5\leqslant l \leqslant A}\nu_i\left(\frac{1}{2}t_z\right) \right]_{TT_z}\right\rangle.
\end{eqnarray}
It depends on the partition of the particles into the $s$-shell, in which the first four nucleons are assigned to the ``alpha-core", the remaining particles are assigned to $p$-shell orbitals. This $s$-shell core is constructed as a simple Slater determinant of spins. Spinors $\chi_i$ and $\nu_i$ specify spin and isospin states of the $p$-shell particles,  and spherical harmonics $Y_{lm_l}$ describe their angular motion around the center of mass of the core, from which they are separated by vectors $\mathbf{r}_{\alpha i}$.

The variational parameters that influence the correlation functions in $\ket{\Psi_J}$, $U_{ij}$ and $U_{ijk}$ are optimized by minimizing the energy expectation value
\begin{equation}
\label{energyexpect}
 E_V = \frac{\bra{\Psi_V}\hat{H}\ket{\Psi_V}}{\braket{\Psi_V|\Psi_V}} \geqslant E,
\end{equation}
which accordingly to the variational principle, is an upper bound to the true ground state energy, $E$. The Metropolis Monte Carlo algorithm is employed to compute the energy expectation value during the optimization process. Due to the nature of Monte Carlo sampling, the variational calculations of observables inherently include statistical uncertainties.

This framework is sufficiently versatile to accommodate unbound states, as demonstrated in previous studies \cite{carlson1984variational,carlson1987microscopic,nollett2007quantum,lynn2016Chiral}. For bound states, VMC methods enforce square-integrable wave functions by ensuring that they decay exponentially at large distances in the $f_{ss}$ correlations and $p$-shell orbitals. However, scattering states present a challenge, as their wave functions extend to infinity, making direct energy minimization problematic. To overcome this, the VMC method can be adapted for scattering by confining the wave function within a spherical box and imposing a boundary condition at the box's edge. In doing so, the VMC scattering problem is transformed into the standard energy-eigenvalue problem. Once this energy is determined, the solution can be matched to the known asymptotic scattering behavior beyond the box. 

A reasonable boundary condition for single-channel scattering is a specified logarithmic derivative $\zeta_c$ of the wave function at the box surface \cite{Thompson_Nunes_2009}, defined by
\begin{equation}
  \label{boundarycond}
\mathbf{\hat{n}}_c\cdot\nabla_{\mathbf{r}_c}\big(\mathbf{r}_c\ket{\Psi}\big) = \zeta_c\mathbf{r}_c\ket{\Psi},
\end{equation}
where the subscript $c$ denotes a scattering channel specified by a division of nucleons into two nuclei and by values of $J,M_J,L,S$. The gradient is calculated in a coordinate system defined by the vector $\mathbf{r}_c$, which separates the centers of mass of the scattering nuclei, with $\mathbf{\hat{n}_c}$ being the outward normal unit vector at the boundary of the box, defined by $r_c = R_0$, where $R_0$ is the radius of the box. For a fixed $R_0$, varying the parameter $\zeta_c$ results in systems with different ground state energies corresponding to scattering at various energies. As for the explicit size of the box it has been found that a radius of $R_0=9\text{ fm}$ is sufficient to be in the asymptotic region of the scattering system \cite{nollett2007quantum}.

The wave function outlined in Eqs. (\ref{varwf})–(\ref{phiA}) can be readily adapted to describe the scattering of a single nucleon by an $s$-shell nucleus. The spin-isospin Slater determinant effectively represents such a $s$-shell nucleus, and the correlations among its nucleons are similar to the $s$-shell portion in Eq.~(\ref{varwf}). We implement the scattering nucleon as a single $p$-shell particle. This particle may carry any orbital angular momentum $L = l$ relative to the center of mass of the nucleus. In general, the coordinate $\mathbf{r}_{\alpha i}$ from Eq.~(\ref{phiA}) must be adjusted to couple the scattered nucleon with the target nucleus to correctly model the $(A-1)+1$ system, however for this work it is the same. The scattering VMC wave function is set by the boundary condition of Eq.~(\ref{boundarycond}) placed on the $p$-shell orbital ($\phi_c$) of Eq.~\ref{phiA}, given by
\begin{equation}
  \label{phiboundarycond}
 (\zeta_c - \frac{1}{R_0})\phi_c(R_0) = \left.\frac{d\phi_c}{dr_c}\right|_{r_c=R_0}.
\end{equation}
Similar to the bound state case, the variational parameters, including the woods-saxon parameters of $\phi_c$ (besides $\zeta_c$), are optimized by minimizing the energy expectation value. In this particular work the VMC $\text{}^5\text{He}$ wave function is constructed by placing the scattered neutron in the $p$-shell and enforcing Eq.~(\ref{phiboundarycond}) on its associated $p$-shell orbital.

%% file: Sections/GFMC.tex
\subsection{\label{subsection:GFMC}Green's function Monte Carlo}
The Green's Function Monte Carlo (GFMC) technique is an advanced method for extracting the true ground-state wave function by refining the initial VMC wave function through the application of the imaginary-time evolution operator \cite{Carlson2015QMC,Lynn:2019rdt,Gandolfi:2020pbj}. This evolution is described by: \begin{equation}\label{gfmcprop} \lim_{\tau \to \infty} e^{-(H-E_{\text{g}})\tau},
\end{equation} where $\tau = i t$, $H$ is the full Hamiltonian of the system under consideration, and $E_{\text{g}}$ is the estimated ground-state energy. In practice the imaginary-time evolution
operator is computed accurately for sufficiently small changes in $\tau$, typically with $\Delta\tau < 0.1$ GeV$^{-1}$, and is usually carried out with a reprojection of the full Hamiltonian, $H^\prime$. This simplified Hamiltonian preserves the isoscalar part in $s$ and $p$ partial waves, as well as the coupling between $\text{}^3D_1$ and $\text{}^3S_1$. Additionally, $H^\prime$ does not directly include higher order terms in the momenta ($\textbf{p}^2,\textbf{L}^2,(\textbf{L}\cdot \textbf{S})^2$,...); these terms are instead dealt perturbatively. Higher-order derivatives can, in principle, be incorporated into the propagator; however, doing so often introduces significant statistical uncertainties \cite{Carlson2015QMC,Lu:2021tab,Curry:2023mkm,Curry:2024gcz}.

The extension to large imaginary time is then achieved through repeated propagation over $\Delta\tau$, resulting in the evolved wave function:
\begin{equation}\label{psitau}  
\Psi(\tau)=\left[e^{-(H^\prime-E_{\text{g}})\Delta\tau}\right]^{N_{\tau}}\Psi_V,
\end{equation}
with the initial condition $\Psi(0) = \Psi_{V}$ and $N_{\tau} = \tau / \Delta \tau$ representing the total number of time steps. The precomputed short-time propagator is evaluated as a Green's function, $G_{\alpha\beta}(\mathbf{R}^\prime,\mathbf{R};\Delta\tau)$ which evolves a spin-isospin vector at position $\mathbf{R}$ in configuration space to a new spin-isospin vector at position $\mathbf{R}^\prime$.  This evolution is determined using a rejection sampling algorithm, as described in \cite{Carlson2015QMC} and is described below. Since $G$ transforms one vector into another, it requires a matrix representation, with each element of the matrix corresponding to
\begin{equation}\label{greens_prop} 
G_{\alpha,\beta}(\textbf{R}^\prime,\textbf{R};\Delta\tau)=\bra{\textbf{R}^\prime,\alpha}e^{-({H}^\prime-E_{\text{g}})\Delta\tau}\ket{\textbf{R},\beta},
\end{equation}
where $\alpha$ and $\beta$ denote spin-isospin indices, omitted in subsequent equations for brevity. Having evaluated the short-time propagator, we can proceed to iteratively construct the wave function for increasing $\tau$ where the final propagation step is represented by:
\begin{equation}\label{psi_n} 
\Psi_{N_\tau}(\textbf{R}_{N_{\tau}}) = \int G(\textbf{R}_{N_{\tau}},\textbf{R}_{N_{\tau}-1};\Delta\tau)\Psi_{N_{\tau}-1}(\textbf{R}_{N_{\tau}-1})d\textbf{R}_{N_{\tau}-1}.
\end{equation}
From here, the entire propagation can be unfolded into a series of repeated integrals over all intermediate configurations, $\mathbf{R}_n$, which are sampled using a Monte Carlo Markov chain approach,
\begin{equation}\label{gfmcchain}
\Psi_{N_\tau}(\textbf{R}_{N_{\tau}})=\int \prod_{n=0}^{N_\tau-1}\big \{G(\textbf{R}_{n+1},\textbf{R}_{n};\Delta\tau)\big \} \Psi_V(\textbf{R}_{0})\prod_{n=0}^{N_\tau-1}d\textbf{R}_{n}.
\end{equation}
These integrals account for the system's evolution from the initial configuration $\textbf{R}_0 $ to the final configuration $\textbf{R}_{N_{\tau}}$. GFMC begins by sampling from the initial variational wave function \( \Psi_V \), with each sample, or "walker," following its own Markov chain path through configuration space as \( \tau \) progresses.

During wave function propagation, the variance in the computed energies can grow exponentially unless corrective measures are applied, a phenomenon known as the fermion sign problem. This issue arises because there is an unphysical bosonic ground state, with significantly lower energy than the fermionic ground state, interferes with the calculation. Sampling individual points cannot enforce the necessary antisymmetry under particle exchange, leading to an increasing contribution from the bosonic state. As a result, when the energy is projected onto the antisymmetric variational wave function (as explained later), the outcome becomes dominated by numerical noise, and the signal is lost. To address this, GFMC uses a constrained path algorithm  \cite{Zhang1997Const,wiringa2000quantum}, ensuring that each walker maintains a positive overlap with the variational wave function at every step: 
\begin{equation}
\braket{\Psi^{\dagger} 
 (\mathbf{R};\tau)|\Psi_V(\mathbf{R})}>0.
\label{cpa}    
\end{equation}
Similarly to the fixed node approximation used to address the sign problem in many electron systems~\cite{Reynolds1982Fixed}, this method relies on the quality of the variational wave function. Fortunately, nuclear VMC is an excellent ansatz~\cite{wiringa91}. However, the constrained path can sometimes lead to convergence on an incorrect energy, either higher or lower than the true value. To address this issue, the constraint is periodically released for a series of steps prior to each energy calculation, helping to minimize errors. For bound states, 10–30 unconstrained steps are typically sufficient to achieve accurate results \cite{Carlson2015QMC}. For unbound states, such as in $\text{}^5\text{He}$, around 80 steps are generally required to avoid introducing a bias of 10–20 keV in the energy evaluation \cite{nollett2007quantum}.

The expectation value of an operator \(\mathcal{O}\) using a propagated GFMC wave function requires only the wave function and the operator itself:
\begin{equation}
\braket{\mathcal{O}}=\bra{\Psi(\tau)}\mathcal{O}\ket{\Psi(\tau)}.
\label{gfmcev}    
\end{equation}
However, if \(\mathcal{O}\) involves derivatives, we encounter the challenge that GFMC provides amplitudes only at discrete points in space, making the computation of derivatives in Eq.~\eqref{gfmcev} impossible. Additionally, the sample points in the left and right propagated wave functions may not align. Despite this, we can express the wave function as:
\begin{equation}
\Psi(\tau)\approx \Psi_V + \delta,
\label{gfmc_perturb}    
\end{equation}
where \(\delta\) is small, and terms of order \(\delta^2\) are negligible. This leads to the commonly used approximation \cite{wiringa2000quantum, Pervin2007Quant, Pieper2009QMC, King2020Weak}:
\begin{equation}
\braket{\mathcal{O}}\approx\braket{\mathcal{O}}_{L}+\braket{\mathcal{O}}_{R}-\braket{\mathcal{O}}_{V}. 
\label{mixedev}    
\end{equation}
Here, subscript $V$ represents the variational expectation value, 
\begin{equation}
\braket{\mathcal{O}}_{V} = \bra{\Psi_V}\mathcal{O}\ket{\Psi_V} , 
\label{vmcev}    
\end{equation}
and $L$ (left) and $R$ (right) indicate the direction propagator acts: 
\begin{eqnarray}
    &&\braket{\mathcal{O}}_{L} = \bra{\Psi(\tau)}\mathcal{O}\ket{\Psi_V}\\\label{leftev}
    &&\braket{\mathcal{O}}_{R} = \bra{\Psi_V}\mathcal{O}\ket{\Psi(\tau)}\label{rightev}.
\end{eqnarray}
For most studies the left and right will be the same, $\braket{\mathcal{O}}_{\text{L}}=\braket{\mathcal{O}}_{\text{R}}$, and these are referred to as ``mixed estimates". The energy expectation value is a special observable because the Hamiltonian commutes with the propagator. For sufficiently large $\tau$, the energy can be computed exactly as: 
\begin{equation}
E=\bra{\Psi(\tau/2)}{H}\ket{\Psi(\tau/2)}=\bra{\Psi(\tau)}{H}\ket{\Psi_V} . 
\label{energyev}    
\end{equation}
Because the simpler ${H}^\prime$ is used to generate the GFMC propagator, the total energy is then computed by the mixed estimate of  ${H}^\prime$ plus the difference $\braket{{H}-{H}^\prime}$ evaluated perturbatively via Eq.~\eqref{mixedev}.


The extension of GFMC to build scattering wave functions follows the same particle-in-a-box framework adopted in the VMC approach. Because GFMC only acts on isolated points in space, we cannot explicitly enforce the logarithmic derivative boundary condition of Eq.~\eqref{boundarycond} nor can we evaluate the wave function outside the box required by Eq.~\eqref{gfmcchain}. Instead, we use the boundary condition, a change of variables, and the method of images (identical to the electrostatics case) to map the integral of Eq.~(\ref{gfmcprop}) to the domain of the box. 

\begin{figure}\centering
\includegraphics[width=8cm]{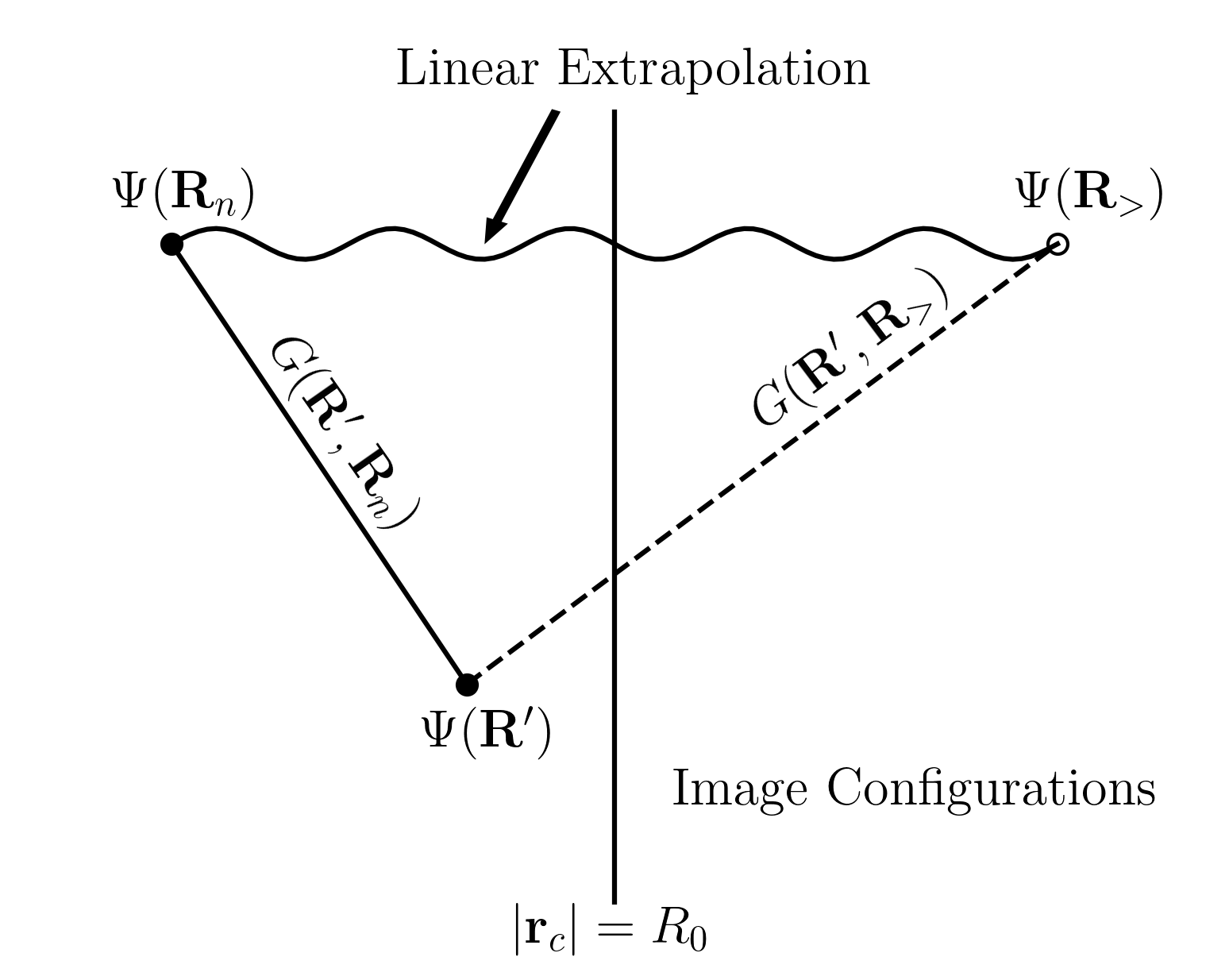}
\caption[]{A simple depiction of how we propagate GFMC scattering states. The GFMC propagation takes the wave function at configuration $\mathbf{R}_n$ and applies the propagator, $G(\mathbf{R}^\prime,\mathbf{R}_n; \Delta\tau)$, to end at the new configuration $\mathbf{R}^\prime$ and $ \Delta\tau$ further in imaginary time. In the unbound case, we are unable to evaluate this propagation outside-the-box ($|\mathbf{r_c}|>R_0$); thus, we use the method of images to determine the contributions to $\Psi(\mathbf{R}^\prime)$ from outside-the-box; represented by the dashed line. However, this requires knowing the wave function outside the box, which in practice is calculated by linear extrapolation using the scattering boundary condition (wavy line), which is only valid near the surface of the box. See the text for the complete expanation. }\label{gfmcscatt}
\end{figure}

To perform this mapping, we start by splitting the GFMC propagation in Eq.~(\ref{gfmcprop}) into contributions from inside- and outside-the-box,
\begin{eqnarray}
&&\Psi_{n+1}(\textbf{R}^\prime)=\int_{|\textbf{r}_{c}|<R_0}G(\textbf{R}^\prime,\textbf{R}_n )\Psi_n(\textbf{R}_n)d\textbf{R}_1d\textbf{R}_2d\textbf{r}_{c}\nonumber\\ &&+\int_{|\textbf{r}_{c}|>R_0}G(\textbf{R}^\prime,\textbf{R}_n )\Psi_n(\textbf{R}_n)d\textbf{R}_1d\textbf{R}_2d\textbf{r}_{>}, 
\label{propinout}    
\end{eqnarray}
where $\mathbf{R}_1$ and $\mathbf{R}_2$ represent the internal cluster coordinates inside the clusters, $\textbf{r}_{c}$ is the separation of cluster centers of mass; $\mathbf{r}_{>}$ is the separation of cluster centers of mass when $|\textbf{r}_{c}|>\text{R}_0$. We then apply a change of variables, $\mathbf{r}_>=(r_>/\text{R}_0)^{2}\mathbf{r_{c}}$  to map the integral of outside-the-box contributions ($|\textbf{r}_{c}|>\text{R}_0$) to inside-the-box, 
\begin{eqnarray}\label{mappedout}   
&&\Psi_{n+1}(\textbf{R}^\prime)=\int_{|\textbf{r}_{c}|<R_0}G(\textbf{R}^\prime,\textbf{R}_n)\Psi_n(\textbf{R}_n)d\textbf{R}_1d\textbf{R}_2d\textbf{r}_{c}\nonumber\\ 
&&+\int_{|\textbf{r}_{c}|<R_0}G(\textbf{R}^\prime,\textbf{R}_>)\left(\frac{r_{>}}{r}\right)^3\Psi_n(\textbf{R}_{>}) d\textbf{R}_1d\textbf{R}_2d\textbf{r}_{c}, 
\end{eqnarray}
where $\mathbf{R}_>$ is the ``image" location outside-the-box implied by the value of $\mathbf{r}_>$. However, the wave function outside the box, $\Psi(\mathbf{R}_>)$, is not directly available. For each sample inside-the-box, the propagation requires the evaluation of the wave function at the mirrored configurations outside the box. Fortunately, the propagator is short-ranged. Consequently, the Green's function that takes the wave function from the image point to the proposed configuration, $G(\mathbf{R}^\prime,\mathbf{R}_>)$, is only significant when ``near'' ($\left|\mathbf{r}-\mathbf{r}_>\right| \leqslant 1 \text{ fm}$) the surface of the box. Thus, a simple linear extrapolation using the log-derivative boundary condition suffices to approximate the wave function at the image configuration. To illustrate the details of this procedure, we have provided Fig.~\ref{gfmcscatt}, a straightforward depiction of the necessary pieces for propagating GFMC scattering states.

The remaining detail is the evaluation of the wavefunction outside the box, $\Psi(\mathbf{R}_>)$. The scattering boundary condition for the GFMC wave function is effectively implemented by using the log-derivative boundary condition in the extrapolation:

\begin{equation}
\Psi_n(\mathbf{R}_{>}) \approx \left[1 + \left(\zeta_c - \frac{1}{\text{R}_0}\right)(\mathbf{R}_{>} - \mathbf{R}_n) \cdot \hat{\mathbf{n}}\right] \Psi_n(\mathbf{R}_n), 
\label{gfmcboundary}
\end{equation}

where \(\zeta_c\) and \(\text{R}_0\) characterize the boundary conditions applied.
Finally, substituting Eq.~(\ref{gfmcboundary}) into Eq.~(\ref{mappedout}) yields to
\begin{equation}
\Psi_{n+1}(\textbf{R}^\prime)=\int_{|\textbf{r}_{c}|<R_0}\left ( G(\textbf{R}^\prime,\textbf{R}_n)+G(\textbf{R}^\prime,\textbf{R}_>)[1+(\zeta_c-\frac{1}{\text{R}_0})(\textbf{R}_{>}-\textbf{R}_n)\cdot \hat{\textbf{n}}]\left(\frac{ r_{>}}{r}\right)^3\right )\Psi_n(\textbf{R}_n)d\textbf{R}_n,
\label{propscatt}    
\end{equation}
which defines the algorithm for propagating GFMC scattering states.  Once Eq.~(\ref{propscatt}) is implemented, the propagation of scattering states in the GFMC method proceeds in a manner similar to the bound-state case.

%% file: Sections/Methods.tex
\section{\label{section:methods}Determination of Scattering Observables}
\input{Sections/Scattering}
\input{Sections/Integral}

\input{Sections/reg}

%% file: Sections/Scattering.tex
In two-cluster scattering, the asymptotic wave function can be described in three common forms, depending on whether the incoming wave is represented by plane, spherical, or standing waves. These forms correspond to the $T$-, $S$-, or $K$-matrix representations, respectively. The $K$-matrix formulation, which is entirely expressed in terms of standing waves, describes the long-range behavior of the wave function as follows:
\begin{equation}
    \label{Kaysmptote}
    \Psi(\text{all } r_c \to \infty)
    = \sum_c  k_c\left(A_c\mathcal{F}_c + B_c\mathcal{G}_c\right),
\end{equation}
where the channel wave number, $k_c$ is given in terms of the channel energy, $E_c$, by $k_c^2 = 2\mu E_c/\hbar^2$, where $\mu$ is the reduced mass.  $\mathcal{F}_c$ and $\mathcal{G}_c$ are the regular and irregular channel-cluster functions, respectively, defined as
\begin{equation}
    \label{Fcompact}
    \mathcal{F}_c \equiv \Psi_{1\otimes2}^c\frac{F_{l_c}(\eta_c,k_cr_c)}{k_cr_c}
\end{equation}
and
\begin{equation}
    \label{Gcompact}
    \mathcal{G}_c \equiv \Psi_{1\otimes2}^c\frac{G_{l_c}(\eta_c,k_cr_c)}{k_cr_c},
\end{equation}
where the Sommerfeld parameter is given by 
\begin{equation}
    \label{sommerfeld}
    \eta = Z_1Z_2e^2\mu/(\hbar^2 k_c).
\end{equation}
$F_l(\eta, \rho)$ and $G_l(\eta, \rho)$ are the usual regular and irregular Coulomb functions \cite{NIST:DLMF}. We define the channel product function of the cluster wave functions $\psi_{1c}$ and $\psi_{2c}$, with specified angular momentum in channel $c$, as
\begin{equation}
    \label{12wf}
    \Psi^c_{1\otimes2} \equiv \mathcal{A}_c\left[\psi_{1c}^{J_{1c}}\otimes\left[\psi_{2c}^{J_{2c}}\otimes Y_{l_c}(\hat{r}_c)\right]_{j_c}\right]_J.
\end{equation}
The operator \(\mathcal{A}_c\) antisymmetrizes the wave function with respect to the partitions of nucleons between the two clusters in channel \(c\), which are described by the wave functions \(\psi_{1c}^{J_{1c}}\) and \(\psi_{2c}^{J_{2c}}\). The angular momentum coupling in Eq.~(\ref{12wf}) organizes the spin and orbital angular momenta (\(J_{2c}\) and \(l_c\)) of cluster 2 to form a total angular momentum \(j_c\). This approach corresponds to "jj coupling," particularly relevant when cluster 2 is a single nucleon, as in the case considered here. The resulting \(j_c\) is then coupled to the angular momentum \(J_{1c}\) of cluster 1, which, for the \(n\text{-}{}^4\text{He}\) system, is an alpha particle with \(J=0\).

The amplitudes $A_c$ and $B_c$ determine all scattering observables, so theoretical calculations aim to find relations among them across all channels. When the amplitudes are written as a column vector, the relation between them is a matrix that predicts scattering outcomes by relating incoming to outgoing amplitudes:
\begin{eqnarray}
&&\mathbf{B} = \hat{K}\mathbf{A}\label{kmat}.
\end{eqnarray}
The amplitude vectors in Eq.~(\ref{kmat}) can, in principle, be read out of any wave function solution and the scattering matrices found by inverting these equations, a process that requires $N_c$ linearly independent wave functions if there are $N_c$ coupled channels.

For $\text{}^5\text{He}$, all cases are single-channel, which means that the matrices are reduced to scalars that can be expressed in terms of a phase shift $\delta$. In single-channel scattering, the relationship between the amplitudes is simplified where $K=B/A=\tan\delta$. We can then directly connect Eqs. (\ref{Kaysmptote}) and  (\ref{boundarycond}), yielding to
\begin{equation}
\label{tandSC}
  \tan\delta = \left .\frac{k\frac{\partial}{\partial \rho}F_{l}(\eta,\rho)-\zeta F_{l}(\eta,\rho)}{\zeta G_{l}(\eta,\rho)-k\frac{\partial}{\partial \rho}G_{l}(\eta,\rho)}\right|_{\rho = kR_0}.
\end{equation}
This approach to calculating single-channel phase shifts is referred to as the \textit{direct method}. The direct method explicitly connects the scattering information to the boundary condition and the energy, eliminating the need to compute the amplitudes \(A\) and \(B\). However, for coupled-channel problems, the computation of surface amplitudes becomes necessary. This challenge motivates the use of the \textit{integral method}, as described below, which provides a systematic framework for addressing coupled-channel problems within QMC calculations.

%% file: Sections/Integral.tex
\subsection{\label{integralsubsec}Integral Method}
Integral methods in scattering provide an efficient framework to understand how particles interact and scatter \cite{lippmann1950variational}. Rather than solving the wave equations point by point, integrals are performed over the interaction region, providing a more manageable approach to extract key information about the scattering process. In this work, we employ the $K$-matrix formalism, defined by the standing wave in Eq.~\eqref{Kaysmptote}. The integral method starts by applying Green's second identity in combination with the Wronskian relation of the irregular and regular coulomb functions \cite{NIST:DLMF} to derive the following normalization for the channel-cluster functions:
\begin{equation}
    \label{wronk}
    \frac{2\mu k_c}{\hbar^2}\left(\Bra{\mathcal{F}_c}H-E\ket{\mathcal{G}_c}
      - \Bra{\mathcal{G}_c}H-E\ket{\mathcal{F}_c} \right)= 1,
\end{equation}
where $H$ is the full many-body Hamiltonian and $E$ is the total energy.  The Dirac bracket here denotes the full contraction of the spin-isospin vector at every point and integration over all independent nucleon coordinates in the center-of-mass frame.  
This expression is non-zero because, in a finite region, the Laplacian operator in the Hamiltonian, $H$, is not Hermitian unless specific boundary conditions are applied~\cite{kievsky2010variational, Romero2011General}. 

Combining the asymptotic form of the scattering wave function in Eq.~\eqref{Kaysmptote} and the normalization of Eq.~\eqref{wronk}, we obtain
\begin{equation}
   \label{fullA}
   A_c = \frac{2\mu}{\hbar^2}\left(\bra{\Psi}H-E\ket{\mathcal{G}_c} - \bra{\mathcal{G}_c}H-E\ket{\Psi} \right) 
\end{equation}
and
\begin{equation}
  \label{fullB}
  B_c = \frac{2\mu}{\hbar^2}\left(\bra{\mathcal{F}_c}H-E\ket{\Psi}
  - \bra{\Psi}H-E\ket{\mathcal{F}_c} \right). 
\end{equation}
A significant challenge in Eq.~\eqref{fullA} arises from the divergence in $G_l$ at $r_c=0$, which leads to computational issues that cannot be handled when performing QMC calculations. One way to remedy this issue is to regularize $G_l$. As long as the regularization of Eq.~\eqref{Gcompact} maintains the relation of Eq.~\eqref{wronk}, the integral method can be carried out. In this work, we follow the same regularization techniques of Ref.~\cite{viviani2020n+} which is expressed as follows:
\begin{eqnarray}
    &&\widetilde{\mathcal{G}}_c = \mathcal{G}_c -\lambda_c(r_c;\gamma)\Psi^c_{1\otimes2}\label{freg}\\
    &&\widetilde{\mathcal{G}}_c(r_c\to 0) = 0\label{freg0}\\
    &&\widetilde{\mathcal{G}}_c(r_c\to R_0) = \mathcal{G}_c \label{fregR0}.
\end{eqnarray}
The explicit form of the regularizer, $\lambda_c$, along with a detailed discussion of its parameters, $\gamma$, is provided in Sec.~\ref{section:reg}. The only modification to the integral relations resulting from the regularization is in  Eq.~\eqref{fullA}, which now becomes
\begin{equation}
\label{FullAreg}
   A_c = \frac{2\mu}{\hbar^2}\left(\bra{\Psi}H-E\ket{\widetilde{\mathcal{G}}_c} - \bra{\widetilde{\mathcal{G}}_c}H-E\ket{\Psi} \right).
\end{equation}
 
The procedure for computing QMC scattering observables via the integral method begins by generating the VMC wave functions as described in Sec.~(\ref{subsection:VMC}), which minimizes Eq.~(\ref{energyexpect}) separately for a scattering state and the individual colliding nuclei.  The optimized wave functions are then propagated to their near exact ground states using the GFMC algorithm outlined in Sec.~(\ref{subsection:GFMC}). This propagation establishes the VMC and GFMC channel energies $E_c$ corresponding to the imposed boundary conditions, Eq.~\eqref{boundarycond}.  That $E_c$ is used with the corresponding wave functions to evaluate the GFMC and VMC estimates of the surface amplitudes, $A_c$ and $B_c$.  Scattering observables are computed from the amplitudes using Eq.~\eqref{kmat}.  One repeats this procedure for many boundary conditions that yield different $E_c$ to obtain results at multiple energies.
\subsection{\label{section:IntVMC}Integral Method in VMC}
In our current computational framework, we do not have full access to the daughters ($1,2$) inside of the cluster product $\Psi^c_{1\otimes2}$. Because of this, we simplify the calculation of Eqs.~(\ref{fullB}) and (\ref{FullAreg}) by assuming that for variational wave functions (denoted with subscript $V$), the daughters ($\alpha$ and $n$) are an eigenstate of their respective Hamiltonian. Specifically, for this work, we assume
\begin{equation}
\label{Valpha_eigen_assume}
\bra{\Psi_V}(H_\alpha - E_{V,\alpha})\ket{\mathcal{W}_{V,c}} \approx 0,
\end{equation}
where $\mathcal{W}=\mathcal{F},\mathcal{G}$, or $\widetilde{\mathcal{G}}$ and $E_{V,\alpha}$ is the VMC energy expectation value for the alpha particle. This assumption should be within Monte Carlo sampling error as long as we use the identical variational parameters of $\psi_{1c}^{J_{1c}}$ for the parameters that constrain the alpha core in $\Psi_V$. This will remain a reasonable assumption for all similar calculations in the $p$-shell given that we constrain the variational parameters of $A-1$ core inside the $A$-body wave function to be identical to the ones of the pure $(A-1)$-body wave function.

We can simplify Eqs.~(\ref{fullB}) and (\ref{FullAreg}) by splitting the Hamiltonian in to a sum of the relative and daughter Hamiltonians, $H=H_\text{rel}+H_\alpha + H_n$, and using Eq.~(\ref{Valpha_eigen_assume}), we obtain
\begin{eqnarray}
   \label{vmcassumefullA_hrel}
   &&A_{V,c} = \frac{2\mu}{\hbar^2}\mathcal{N}_V\left(\bra{\Psi_V}H_\text{rel}-E_{\text{rel}} \ket{\widetilde{\mathcal{G}}_{V,c}}-\bra{\widetilde{\mathcal{G}}_{V,c}}H-E_V\ket{\Psi_V}\right),
\end{eqnarray}
and
\begin{equation}
  \label{vmcassumefullB_hrel}
  B_{V,c} = \frac{2\mu}{\hbar^2}\mathcal{N}_V\left(\bra{\mathcal{F}_{V,c}}H-E_V\ket{\Psi_V}
  - \bra{\Psi_V}H_\text{rel}-E_{\text{rel}}\ket{\mathcal{F}_{V,c}} \right), 
\end{equation}
with the norm being
\begin{equation}
    \mathcal{N}_V=(\braket{\Psi_V|\Psi_V}\braket{\psi_{1c}^{J_{1c}}|\psi_{1c}^{J_{1c}}})^{-1/2},
\end{equation}
and the relative Hamiltonian defined as
\begin{equation}
  \label{hrel}
  H_\text{rel} = V_\text{rel}^c -\frac{\hbar^2}{2\mu}\nabla^2_{r_c}.
\end{equation}
Here $V_\text{rel}^c$ is the full relative potential between the daughters. For this work, it is the interaction between the neutron and all particles within the alpha.  We have separated the total energy, $E_V$, into $E_V=E_{\text{rel}}+E_{V,\alpha}$ and disregard the energy associated with the neutron because $(H_n-E_n)\ket{\psi_n}$ is trivially zero in its internal frame, where $\psi_n$ is a spinor with $(S,T)=(1/2, 1/2)$. It is useful to note that the relative energy, $E_{\text{rel}}$, is an input parameter to the cluster-product functions $\mathcal{F}$ and $\mathcal{G}$. Additionally, we can further simplify the surface amplitudes by noticing that the relative kinetic energy only acts on  the Coulomb functions (with a factor of $1/r_c$), and the spherical harmonic inside the factorized wave function, $\Psi_{1\otimes2}$. Using the definition of the Coulomb wave functions and the $L^2$ form of the Laplacian, one finds that the relative Hamiltonian reduces to, 
\begin{equation}
  \label{F_hrel_simp}
    H_\text{rel}\ket{\mathcal{F}_{c}} = (V_\text{rel}^c -V_\mathcal{C}^c)\ket{\mathcal{F}_{c}}.
\end{equation}
Here $V_\mathcal{C}^c$ is the Coulomb potential between the daughters, which is zero for this system. As for the regularized Coulomb function we find a similar reduction, 
\begin{equation}
  \label{greg_hrel_simp}
    H_\text{rel}\ket{\widetilde{\mathcal{G}}_{c}} = (V_\text{rel}^c -V_\mathcal{C}^c)\ket{\mathcal{G}_{c}} - H_\text{rel}\ket{\lambda_c\Psi^c_{1\otimes2}}.
\end{equation}
The simplification of $H_\text{rel}$ applies to all cluster-product functions not just the variational ones. We can then apply Eq.~\eqref{greg_hrel_simp} to Eq.~\eqref{vmcassumefullA_hrel} and find: 
\begin{eqnarray}
   \label{vmcassumefullA}
  &&A_{V,c} = \frac{2\mu}{\hbar^2}\mathcal{N}_V\left(\bra{\Psi_V}V_\text{rel}^c-V_\mathcal{C}^c\ket{\mathcal{G}_{V,c}} -\bra{\mathcal{G}_{V,c}}H-E_V\ket{\Psi_V} - \Lambda_{V,c}\right),
\end{eqnarray}
where we have condensed the regularization integrals into
\begin{equation}
   \label{ireg_full}
  \Lambda_{c}(\gamma) =  \bra{\Psi}V_\text{rel}^c -\frac{\hbar^2}{2\mu}\nabla^2_{r_c}-E_{\text{rel}}\ket{\lambda(r_c,\gamma)\Psi^c_{1\otimes2}}-\bra{\lambda_c(r_c,\gamma)\Psi^c_{1\otimes2}}H-E\ket{\Psi}.
\end{equation}
It is useful to analytically compute the relative Laplacian, which results in the following expression:
\begin{equation}
   \label{ireg_full_simp}
  \Lambda_{c}(\gamma) =  \bra{\Psi}V_\text{rel}^c +\frac{\hbar^2}{2\mu}(\frac{l(l+1)}{r_c^2}-\frac{2}{r_c}\frac{\lambda^\prime_c}{\lambda_c}-\frac{\lambda^{\prime\prime}_c}{\lambda_c})-E_{\text{rel}}\ket{\lambda_c\Psi^c_{1\otimes2}}-\bra{\lambda_c\Psi^c_{1\otimes2}}H-E\ket{\Psi},
\end{equation}
where the prime (\('\)) represents a derivative with respect to the cluster separation \(r_c\). Lastly, we can also apply the simplification of Eq.~\eqref{F_hrel_simp} to Eq.~\eqref{vmcassumefullB_hrel},
\begin{equation}
  \label{vmcassumefullB}
  B_{V,c} = \frac{2\mu}{\hbar^2}\mathcal{N}_V\left(\bra{\mathcal{F}_{V,c}}H-E_V\ket{\Psi_V}
  - \bra{\Psi_V}V_\text{rel}^c-V_\mathcal{C}^c\ket{\mathcal{F}_{V,c}} \right). 
\end{equation}
These integrals are the ones we compute to get the phase shifts, and  we report the calculation of Eqs.~(\ref{vmcassumefullA}) and (\ref{vmcassumefullB}) in Sec.~\ref{section:results} as phase shifts, denoting them as ``Integral VMC."
\subsection{\label{section:intGFMC}Integral Method in GFMC}
The objective of the integral method in GFMC is to compute the imaginary time propagated expectation values for the surface amplitudes, $A$ and $B$, as described in the previous section. In principle we would like to compute,  
\begin{equation}
\label{FullAreg_gfmc}
   A_c(\tau) \equiv \frac{2\mu}{\hbar^2}\mathcal{N(\tau)}\left(\bra{\Psi(\tau)}H-E\ket{\widetilde{\mathcal{G}}_c(\tau)} - \bra{\widetilde{\mathcal{G}}_c(\tau)}H-E\ket{\Psi(\tau)} \right),
\end{equation}
and 
\begin{equation}
  \label{fullB_gfmc}
  B_c(\tau) \equiv \frac{2\mu}{\hbar^2}\mathcal{N(\tau)}\left(\bra{\mathcal{F}_c(\tau)}H-E\ket{\Psi(\tau)}
  - \bra{\Psi(\tau)}H-E\ket{\mathcal{F}_c(\tau)} \right), 
\end{equation}
with normalization, $\mathcal{N}(\tau)=(\braket{\psi_{1c}^{J_{1c}}(\tau)|\psi_{1c}^{J_{1c}}(\tau)}\braket{\Psi(\tau)|\Psi(\tau)})^{-1/2}$. Here we have introduced the propagated channel-cluster functions $\mathcal{F}_c(\tau)$ and $\widetilde{\mathcal{G}}_c(\tau)$, which differ from Eq.~(\ref{Fcompact}) and Eq.~(\ref{freg}) only by propagation of the alpha particle defined in Eq.~(\ref{12wf}), 
\begin{equation}
    \label{12wfprop}
    \Psi^c_{1\otimes2}(\tau) \equiv \mathcal{A}_c\left[\psi_{1c}^{J_{1c}}(\tau)\otimes\left[\psi_{2c}^{J_{2c}}\otimes Y_{l_c}(\hat{r}_c)\right]_{j_c}\right]_J.
\end{equation}
However, in practice, we must compute mixed estimates, Eq.~\eqref{mixedev}, of the surface amplitudes. The mixed estimates for the surface amplitudes are a special kind, as the functions to the left and right of the operator are not the same. These mixed estimates are referred to as ``off-diagonal" mixed estimates \cite{Pervin2007Quant}. The additional difficulty for off-diagonal mixed estimates of the Eqs.~(\ref{FullAreg_gfmc}) and (\ref{fullB_gfmc}) is that the propagated walk for the channel-cluster function only contains information about the alpha-particle, and does not know about the scattering neutron. To perform the calculation, we must assign a probability distribution $\rho(r_c)$ for the cluster separation, $r_c$. In principle, any distribution will suffice; however, this choice significantly affects the Monte Carlo statistical variance. To minimize the variance, $\left|\psi_{1c}^{J_{1c}} \cdot \rho(r_c)\right|^2$ should closely approximate the sampling density of the $A$-body wave function. Moreover, to preserve the normalizations, we require that $\int_{r_c<R_0}\rho^2(r_c)dr_c =1 $. A rigorous study on the impacts of this sampling distribution on these kinds of mixed estimates was carried out in Ref.~\cite{brida2011quantum}. That study found that for bound states, either a Gaussian or a Woods-Saxon function works well for $\rho(r_c)$. However, for scattering calculations, we have found that an exponential decay function, 
\begin{equation}\label{expsample}
   \rho(r_c) = \sqrt{2}e^{-r_{c}},
\end{equation}
reduces sampling variances by increasing the sampling where all the particles are strongly interacting, meaning where $V_\text{rel}$ is large. 

To calculate Eqs.~(\ref{FullAreg_gfmc}) and (\ref{fullB_gfmc}) one can, neglecting terms of order $[\Psi(\tau)-\Psi_V]^2$ and $[\psi_{1c}^{J_{1c}}(\tau)-\psi_{1c,V}^{J_{1c}}]^2$, find a similar result to Eq.~(12) in Ref.~\cite{brida2011quantum}:
\begin{equation}
   \label{Aoftau}
   A_c(\tau) \approx A_c(\Psi(\tau))+A_c(\psi_{1c}^{J_{1c}}(\tau))-A_{c,V},
\end{equation}
and
\begin{equation}
   \label{Boftau}
   B_c(\tau) \approx B_c(\Psi(\tau))+B_c(\psi_{1c}^{J_{1c}}(\tau))-B_{c,V}.
\end{equation}
The subscript $V$ denotes the variational surface amplitude defined in Eq.~(\ref{vmcassumefullA}) and Eq.~(\ref{vmcassumefullB}). Here we have introduced the off mixed estimate where we have propagated the full $A$-body scattering wave function, $A_c(\Psi(\tau),B_c(\Psi(\tau))$, and where we have propagated the $(A-1)$ daughter wave function $A_c(\psi_{1c}^{J_{1c}}),B_c(\psi_{1c}^{J_{1c}})$. The propagated full $A$-body case are defined as,
\begin{equation}
   \label{abody_A}
   A_c(\Psi(\tau)) \equiv  \frac{2\mu}{\hbar^2}\frac{\sqrt{\mathcal{N}}}{(\braket{\Psi_V|\Psi(\tau)}}\left(\bra{\Psi(\tau)}V_\text{rel}^c-V_\mathcal{C}^c \ket{\mathcal{G}_{c,V}}-\Lambda_c(\Psi(\tau))\right),
\end{equation}
where,
\begin{equation}
   \label{abody_A_lambda}
   \Lambda_c(\Psi(\tau)) = \bra{\Psi(\tau)}H_\text{rel}-E_\text{rel}\ket{\lambda_c(r_c;\gamma)\Psi_{1\otimes2}^c}.
\end{equation}
Similarly,
\begin{equation}
   \label{abody_B}
   B_c(\Psi(\tau)) \equiv  \frac{2\mu}{\hbar^2}\frac{\sqrt{\mathcal{N}}}{\braket{\Psi_V|\Psi(\tau)}}\bra{\Psi(\tau)}V_\text{rel}^c-V_\mathcal{C}^c \ket{\mathcal{F}_{c,V}},
\end{equation}
where we have introduced the the self-normalization factor, 
\begin{equation}
   \label{selfnorm}
   \mathcal{N}\equiv \frac{\braket{\Psi_V|\Psi_V}}{\braket{\psi_{1c}^{J_{1c}}|\psi_{1c}^{J_{1c}}}}.
\end{equation}
For Eqs.~(\ref{abody_A}) and (\ref{abody_B}) we have made use of the fact that,
\begin{equation}
   \label{propIsTrue}
   (H-E)\ket{\Psi(\tau)}=0,
\end{equation}
and
\begin{equation}
   \label{propIsTrue_alpha}
   \bra{\Psi(\tau)}(H_\alpha-E_\alpha)\ket{\mathcal{W}_{c,V}}=0,
\end{equation}
where $\mathcal{W}=\mathcal{F},\mathcal{G},$ or $\widetilde{\mathcal{G}}$. These are the properties of the true ground state of the wave function. Similarly, we make use of the property,
\begin{equation}
\label{alpha_eigen_assume}
(H_\alpha - E_\alpha)\ket{\mathcal{W}_c(\tau)} = 0,
\end{equation}
for the amplitudes when we have propagated the alpha particle. Those amplitudes are defined as,
\begin{equation}
   \label{am1body_A}
   A_c(\psi_{1c}^{J_{1c}}(\tau)) \equiv  \frac{2\mu}{\hbar^2}\frac{\sqrt{1/       
                            \mathcal{N}}}{\braket{\psi_{1c}^{J_{1c}}(\tau)|\psi_{1c}^{J_{1c}}}}
                            \left(\bra{\Psi_V}V_\text{rel}^c-V_\mathcal{C}^c \ket{\mathcal{G}_c(\tau)}
                            -\bra{\mathcal{G}_c(\tau)}H-E\ket{\Psi_V}-\Lambda_c(\psi_{1c}^{J_{1c}}(\tau))\right),
\end{equation}
where the regularization integrals are, 
\begin{equation}
   \label{am1body_lambda}
   \Lambda_c(\psi_{1c}^{J_{1c}}(\tau)) \equiv  \bra{\Psi}H_\text{rel}-E_{\text{rel}}\ket{\lambda_c\Psi^c_{1\otimes2}(\tau)}-\bra{\lambda_c\Psi^c_{1\otimes2}(\tau)}H-E\ket{\Psi},
\end{equation}
and 
\begin{equation}
   \label{am1body_B}
   B_c(\psi_{1c}^{J_{1c}}(\tau)) \equiv  \frac{2\mu}{\hbar^2}\frac{\sqrt{1/       
                            \mathcal{N}}}{\braket{\psi_{1c}^{J_{1c}}(\tau)|\psi_{1c}^{J_{1c}}}}
                            \left(\bra{\mathcal{F}_c(\tau)}H-E\ket{\Psi_V}-\bra{\Psi_V}V_\text{rel}^c-V_\mathcal{C}^c \ket{\mathcal{F}_c(\tau)}\right).
\end{equation}
Lastly, there is additional care necessary when computing the variational part of the Eq.~\eqref{Aoftau} and Eq.~\eqref{Boftau}. Specifically, the detail comes into play from our assumption in Eq.~\eqref{Valpha_eigen_assume}, that is for the true energy,  $\bra{\Psi_V}(H_\alpha - E_\alpha)\ket{\mathcal{W}_{V,c}} \ne 0$, where $\mathcal{W}=\mathcal{F},\mathcal{G},$ or $\widetilde{\mathcal{G}}$. However, we can instead write,
\begin{equation}
\label{Valpha_eigen_mixed}
\bra{\Psi_V}(H_\alpha - E_\alpha + E_{\alpha,V} - E_{\alpha,V})\ket{\mathcal{W}_{V,c}} \approx \bra{\Psi_V}\Delta_\alpha\ket{\mathcal{W}_{V,c}},
\end{equation}
where $\Delta_\alpha$ is the difference between the true and variational energy of the alpha particle;
\begin{equation}
\label{delta_alpha}
\Delta_\alpha = E_\alpha(\tau=0)-E_\alpha(\tau).
\end{equation}
The modification of Eqs.~(\ref{vmcassumefullA}) and (\ref{vmcassumefullB}) for Eqs.~(\ref{Aoftau}) and (\ref{Boftau}), respectively, is a simple replacement of $E_V$ with $E+\Delta_\alpha$. This adjustment can be understood as using the true relative energy of the system but with the variational energy of the daughters. This is because there is only variational wave functions but the relative energy must still be exact.

The final task is to take the limit as imaginary time approaches infinity. In practice, this involves sampling until the energy stabilizes. Regardless of the chosen boundary condition, the energy will converge to the ground state energy of the system, as determined by the boundary condition. We present the off-diagonal mixed estimates from Eqs.~(\ref{Aoftau}) and (\ref{Boftau}) as phase shifts in Sec.~\ref{section:results}, labeled as ``Integral GFMC".

%% file: Sections/reg.tex
\section{Regularization\label{section:reg}}
In order to ensure a correct evaluation of Eq.~(\ref{FullAreg}), we need to verify that our regularizer does not bias our phase shifts. To be precise we must guarantee that the normalization: 
\begin{equation}
    \label{normreg}
    \frac{2\mu k_c}{\hbar^2}\left(\Bra{\mathcal{F}_c}H-E\ket{\widetilde{\mathcal{G}_c}}
      - \bra{\widetilde{\mathcal{G}_c}}H-E\ket{\mathcal{F}_c} \right)= 1,
\end{equation}
holds at the surface of the box for our choice of regularizer. The explicit form of our regularizer is equivalent to Eq. (6) in Ref.~\cite{viviani2020n+} (the regularizer used in that work). Specifically, for neutron scattering the Sommerfield parameter, Eq.~\eqref{sommerfeld} is zero; $\eta=0$. For $L=0$ channels, our regularizer is
\begin{equation}\label{reg_l0}
    \displaystyle \lambda_0(r;\gamma)=\frac{e^{-\gamma r}}{kr}(1+\frac{\gamma^2 + k^2}{2\gamma}r),
\end{equation}
and for $L=1$, 
\begin{equation}\label{regl1}
    \displaystyle \lambda_1(r;\gamma)=\frac{e^{-\gamma r}}{k^2r^2}(1+\gamma r + \frac{\gamma^2+k^2}{2}r^2),
\end{equation}
where $k$ is the wave-number, and $\gamma$ is a parameter. In general, we should find that for a choice of $\gamma$ that satisfies the normalization of Eq.~(\ref{normreg}), the regularization integrals, $\Lambda_c$, defined in Eq.~(\ref{abody_A_lambda}), should converge. To test this idea, we can analyze the phase shifts for a specific scattering channel at a given center-of-mass energy, and study how these phase shifts depend on the regularization parameter $\gamma$. Using the phase shifts in our analysis instead of the unnormalized value of the regularization integrals ($\Lambda$), we can better understand how the regularization parameter ($\gamma$) influences our results by comparison with the direct method.

\begin{figure}\centering
\includegraphics[width=12cm]{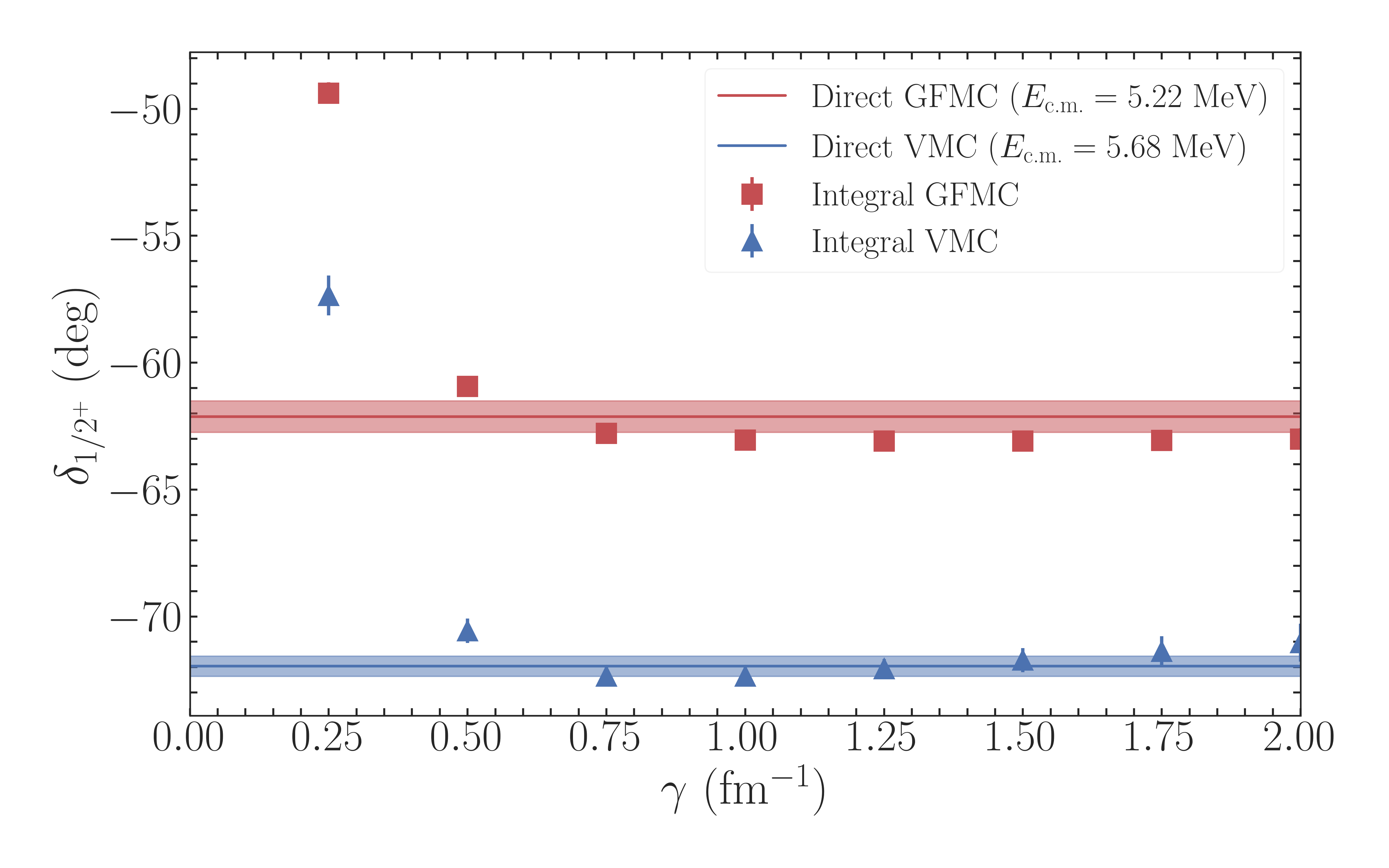}
\caption[]{$J^\pi=1/2^+$ phase shifts (in degrees) for $n+\text{}^4\text{He}$ as a function of regularizer parameter $\gamma$. The integral method (GFMC: red squares, VMC: blue triangles) should agree with the direct method (solid line with error band, GFMC: red, VMC: blue) and not be impacted by the regularization parameter. The stabilization of the phase shifts for $\gamma > 0.75$ (fm$\text{}^{-1}$) indicates that the regularization integrals $\Lambda_c$ converge and do not bias our calculations.}\label{shiftsreg1hp2}
\end{figure}

We begin our analysis by calculating Eqs.~(\ref{vmcassumefullA}) and (\ref{vmcassumefullB}) for VMC and the mixed estimates of Eqs.~(\ref{Aoftau}) and (\ref{Boftau}) for GFMC. We benchmark these results with direct method phase shifts calculated using Eq.~\eqref{tandSC}. However, the more important test is to inspect the rate of change of the phase shifts with respect to $\gamma$; this should be zero when Eq.~(\ref{normreg}) is satisfied. We evaluate the phase shifts as a function of the regularization parameter, as seen for the $s$-wave in Fig.~\ref{shiftsreg1hp2}. We compare the Integral methods (GFMC: red squares, VMC: blue triangles) to the direct phase shift (solid line with error band, GFMC: red, VMC: blue) for a single energy (GFMC: $E_{\text{c.m.}}=5.22$ MeV, VMC: $E_{\text{c.m.}}=5.68$ MeV) in the $1/2^+$ channel. In this case, the integral methods show no impact from the choice of $\gamma$ above $0.75$ (fm$\text{}^{-1}$). This is the ideal result. 

\begin{figure}\centering
\includegraphics[width=12cm]{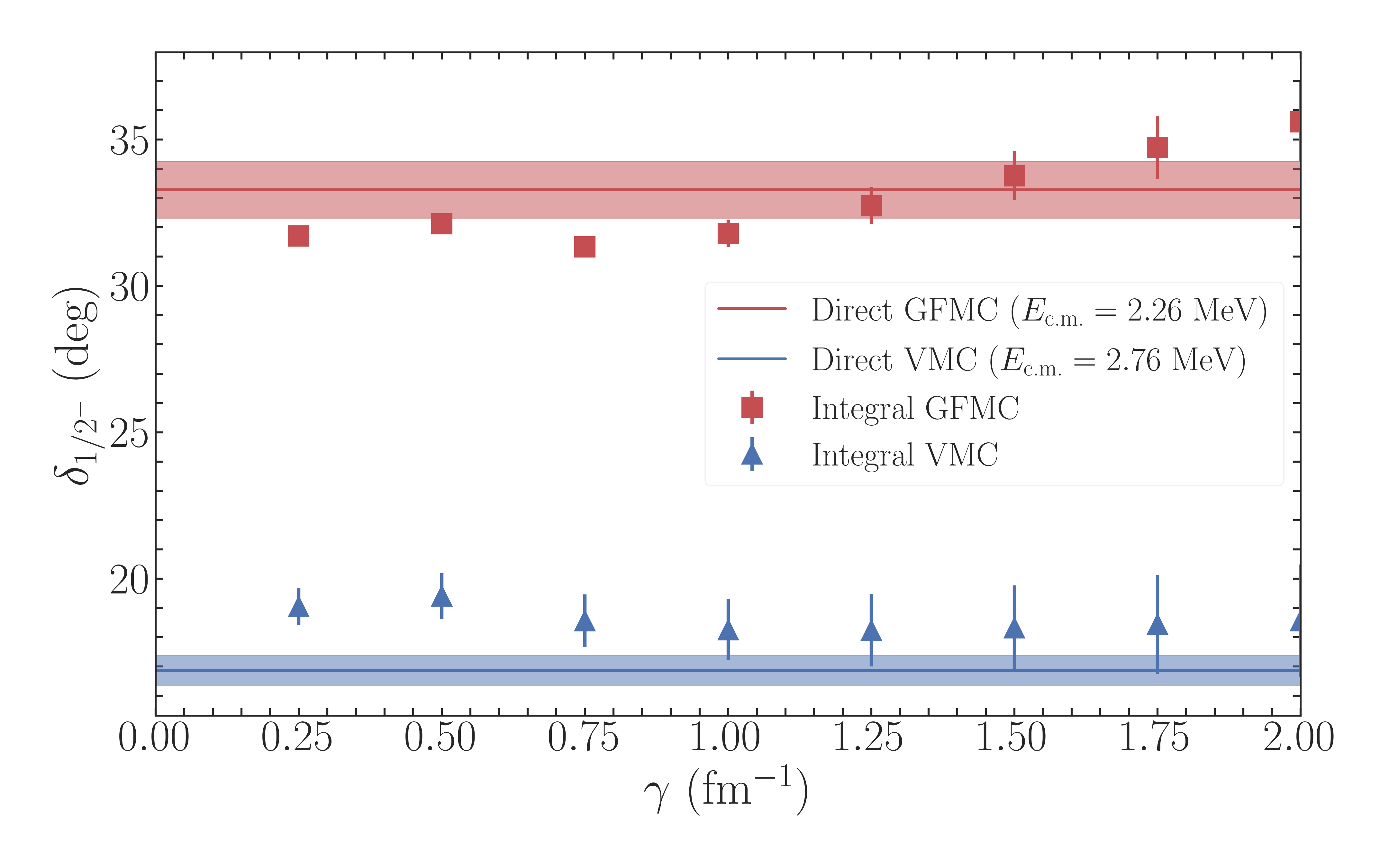}
\caption[]{$J^\pi=1/2^-$ phase shifts (in degrees) for $n+\text{}^4\text{He}$ as a function of regularizer parameter $\gamma$. Labels and methods similar to those in Fig.~\ref{shiftsreg1hp2}}\label{shiftsreg1hm}
\end{figure}

Unfortunately, we see a slight bias for the GFMC integral method in both $p$-wave wave channels. The analysis of the $1/2^-$ channel is identical to that of the $3/2^-$; thus, we only report on the $1/2^-$ channel for brevity. In Fig.~\ref{shiftsreg1hm}, we compute the phase shifts as a function of the regularization parameter for the $1/2^-$ channel at $E_{\text{c.m.}}=2.26$ MeV (and at $E_{\text{c.m.}}=2.76$ MeV for VMC). We find that for $0.25 < \gamma < 1.0$ fm$\text{}^{-1}$, there is a region of stability that does not bias the phase shifts. However, we must not exceed a computational upper bound for $\gamma$ for our numerical integrals inside the box (radius 9 fm). We also see that the VMC central value is constant, but for both integral methods, the sampling error grows directly with $\gamma$. The error growth is understood that as we increase $\gamma$, we are getting closer to a regularizer that looks like a step function, which will, in theory, be better but, in practice, cause numerical difficulty. 

The computational problem is that as $\gamma$ increases, we reduce the radius ($r_c$) that $\Lambda(r_c)$ is non-zero. For $\Lambda$ to be computed accurately means sampling a volume significantly smaller than our box's overall volume, which is not assessable with the current importance sampling. The net result is that only a few sampling points fall within the importance region for $\Lambda$ with $\gamma > 1\text{ }\text{fm}^{-1}$, which increases the variance and causes a loss of information due to numerical cutoffs. It is likely due to this numerical instability that we see the bias in the integral GFMC phase shifts. Therefore, we find that $\gamma=1.0$ fm$\text{}^{-1}$ is a reasonable trade-off for computational and physical accuracy. All integral method results presented in Sec.~\ref{section:results} are calculated using this choice of $\gamma$.

%% file: Sections/Results.tex
\section{\label{section:results}Results}
\input{Sections/Energy}
\input{Sections/Shifts}

%% file: Sections/Energy.tex
\begin{figure}\centering
\includegraphics[width=12cm]{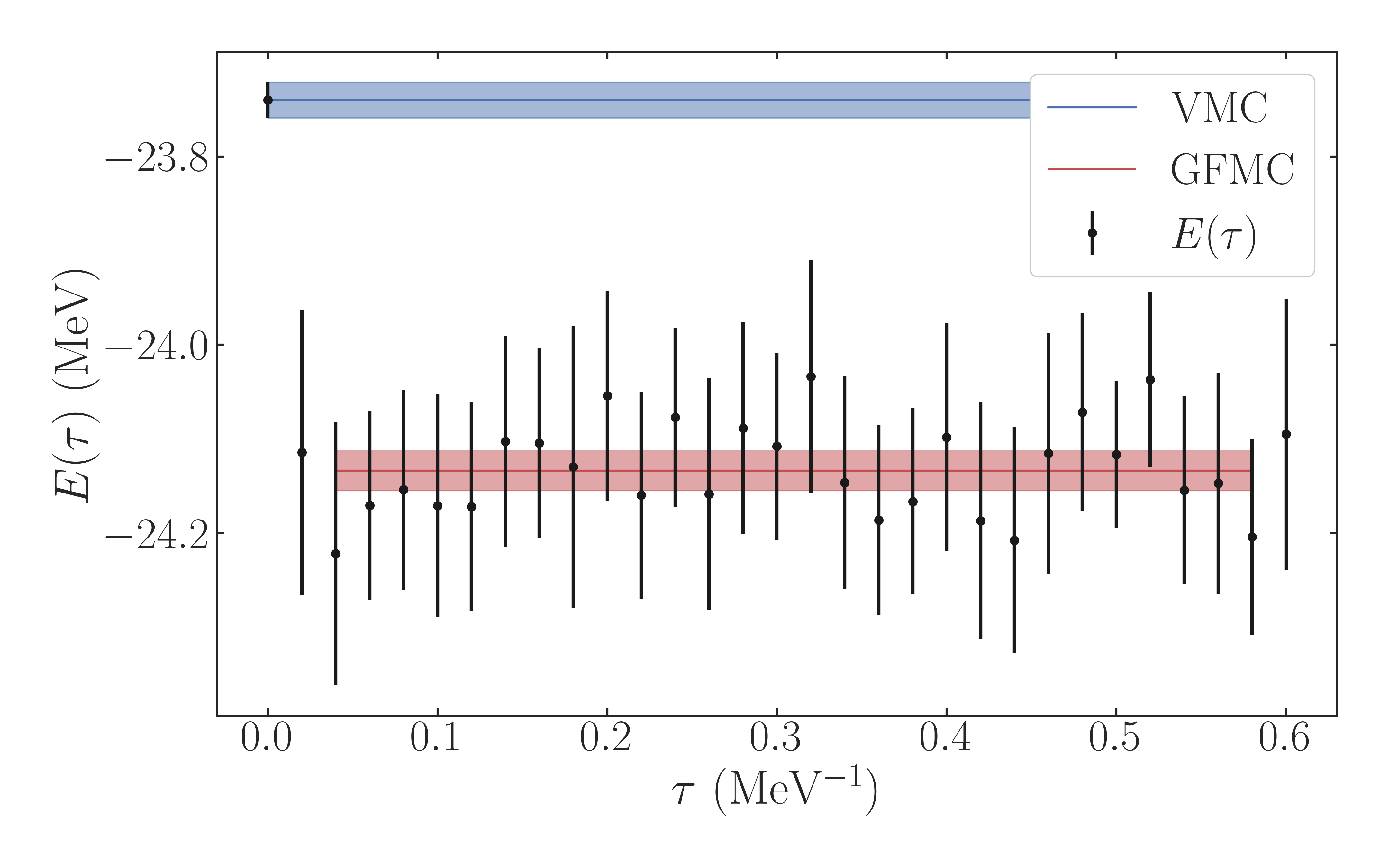}
\caption[]{Ground state $\text{}^4\text{He}$ quantum Monte Carlo energy expectation values (in MeV) as a function of imaginary time ($\tau$) for the AV18 nuclear interaction. The wave function yielding the variational Monte Carlo estimate (solid blue line with error band) and the $\tau=0$ value for Green's function Monte Carlo constitute the initial starting point for GFMC. Energies are computed at intervals of 40$\Delta\tau$ from walkers allowed to propagate unconstrained for 25 steps before the evaluation. Each energy expectation value shown as a black dot is an average over Monte Carlo samples at the indicated $\tau$. The horizontal red line and error band are the average of those expectation values from $\tau=0.04 \text{ to } 0.58 \text{ MeV}^{-1}$. The GFMC method quickly removes high-energy contamination from the initial VMC wave function.}\label{he4energy}
\end{figure}

We begin our discussion of the results by examining the GFMC energy expectation values as a function of imaginary time $\tau$, using a time step of $\Delta\tau=0.0005 \text{ MeV}^{-1}$, to illustrate the stabilization of the propagated energy. We first focus on the energy of the alpha particle, which establishes the threshold. In Fig.~\ref{he4energy}, we present the bound-state energy of \(^4\text{He}\) as a function of \( \tau \), computed using the AV18 potential. As expected, the propagated energies (black circles with error bars) rapidly suppress the high-energy contamination present in the VMC energy (blue band) and converge to a stable value at large \( \tau \), represented by the average indicated by the red band. The VMC alpha binding energy we computed with this potential is $-23.75(1)$ MeV compared to the GFMC result of $-24.12(2)$ MeV, which is in excellent agreement with previously benchmarked QMC results \cite{Pieper2009QMC} and recent hyperspherical harmonic method calculations \cite{marcucci2020hyperspherical}.

\begin{figure}\centering
\includegraphics[width=12cm]{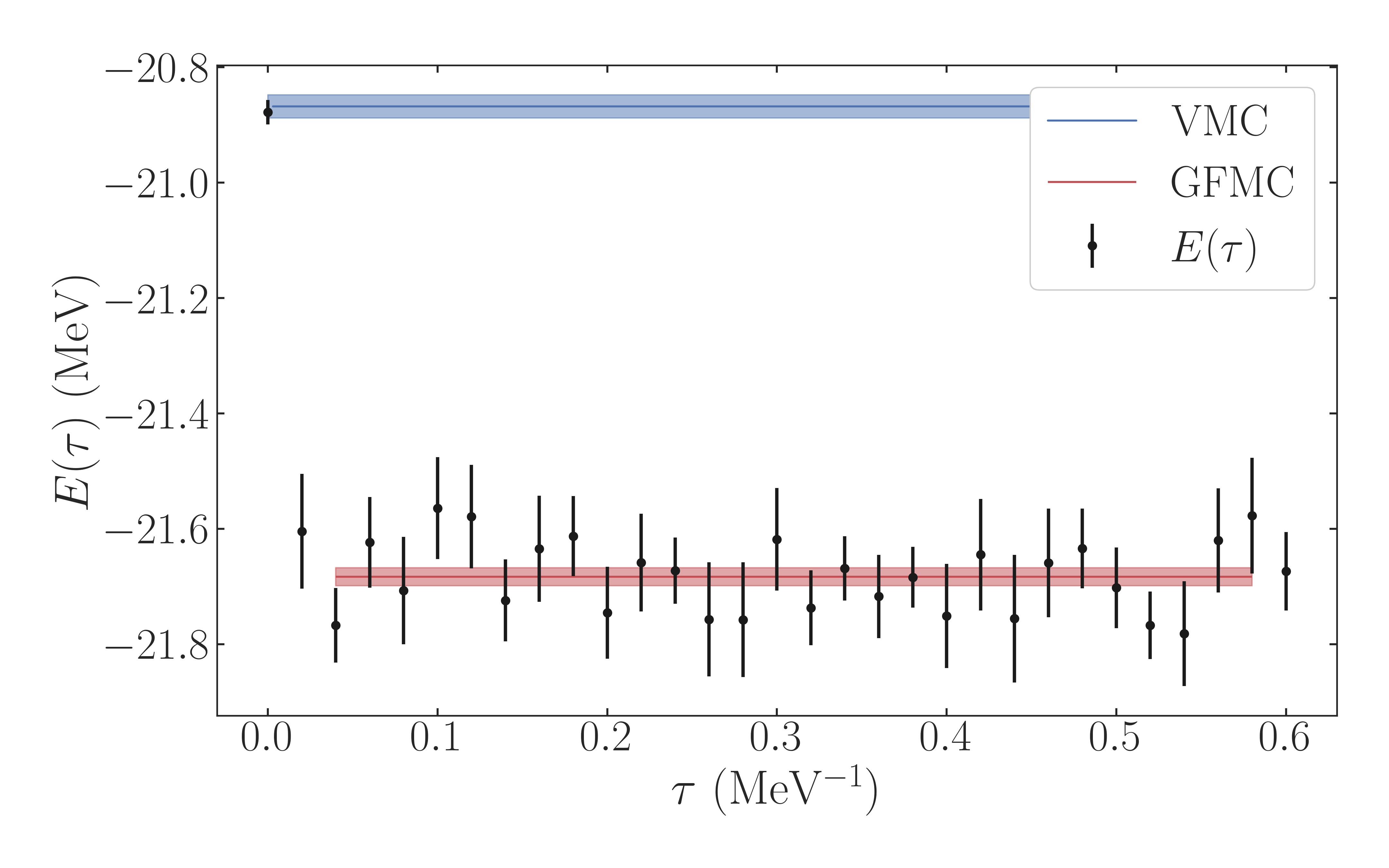}
\caption[]{GFMC energy propagation of a scattering state in the $\text{}^5\text{He}$ $1/2+$ system for AV18.  The boundary condition is $\zeta=-0.13 \text{ fm}^{-1}$ at 9 fm, and symbols are as in Fig.~\ref{he4energy}. In this case, energies were computed for every 40 steps using walkers that propagated 80 final steps without constraint.}\label{1halfpE}
\end{figure}

Next, we calculate the unbound \(^5\text{He}\), low-energy scattering of neutrons from alpha particles is dominated by three uncoupled partial waves. The channels are characterized by total angular momentum and parity \( J^{\pi} = 1/2^{+}, 1/2^{-}, \) and \( 3/2^{-} \). The positive parity channel corresponds to an $s$-wave (\( L = 0 \)), while the negative parity channels are $p$-waves (\( L = 1 \)). The process of computing energies is identical in every channel; therefore, we illustrate the procedure using the $1/2^{+}$ channel in Fig.~\ref{1halfpE}. We begin by optimizing a VMC wave function for a specific boundary condition ($\zeta=-0.13 \text{ fm}^{-1}$ at 9 fm) and then propagating that wave function according to Eq.~(\ref{propscatt}). We then repeat the propagation, updating a VMC parameter associated with the relative energy to have a self-consistent value between the trial wavefunction and the GFMC energy. Similarly to the bound case, the energy expectation value depends on imaginary time ($\tau$), and we illustrate this dependence for the representative $1/2+$ state in Fig.~\ref{1halfpE}.

\begin{figure}\centering
\includegraphics[width=12cm]{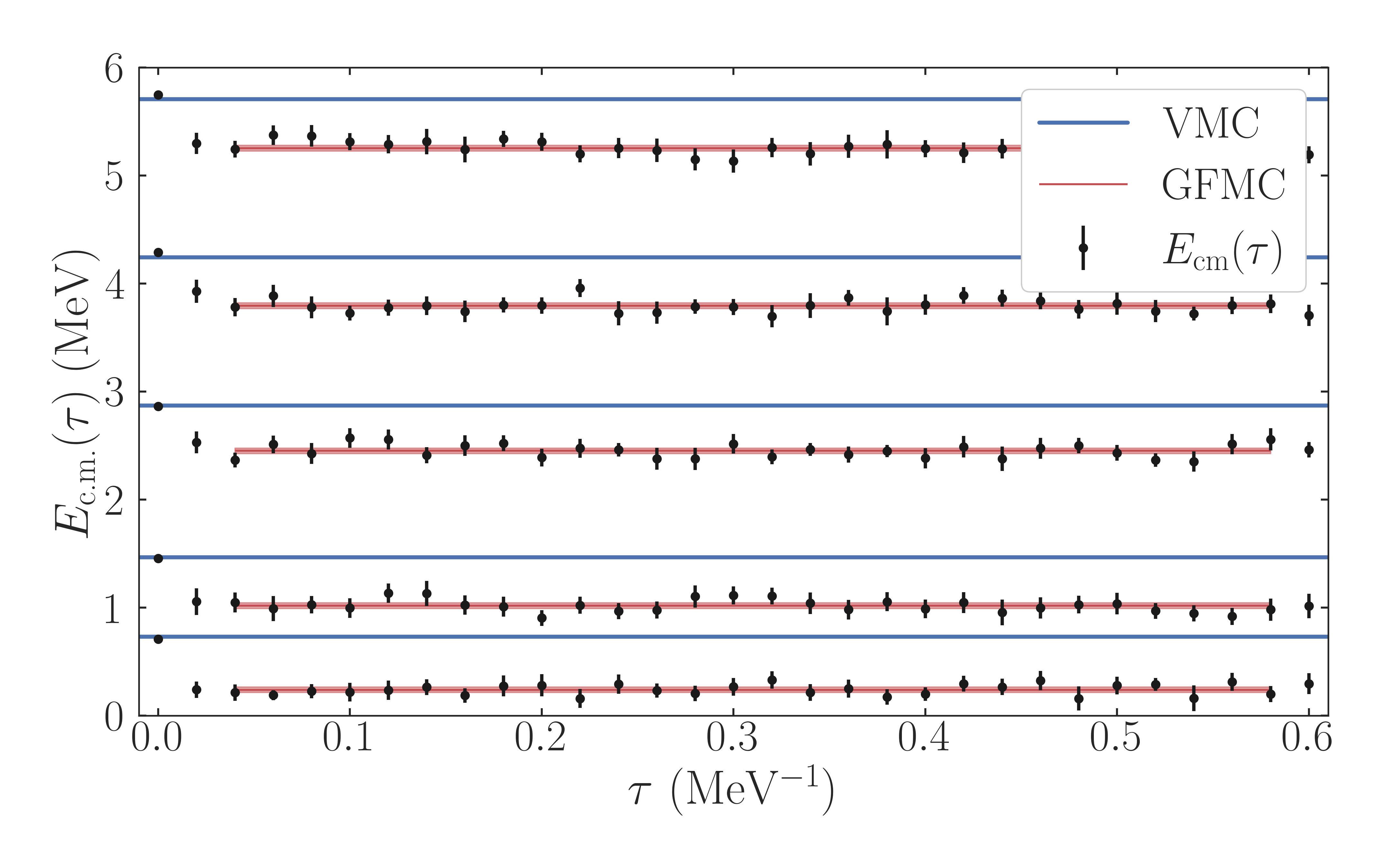}
\caption[]{Energies relative to threshold of $n+\text{}^4\text{He}$ scattering states with $J^\pi=1/2^+$ (in MeV) as functions of imaginary time ($\tau$) for the AV18 nuclear interaction and various boundary conditions. Each choice of $\zeta$ has its own set of energy values. Symbols and methods are as in Figs.~\ref{he4energy} and \ref{1halfpE}. In each case, GFMC quickly settles on a stable energy corresponding to that boundary condition.}\label{1halfpEall}
\end{figure}

The propagated energy of the unbound state stabilizes in a manner similar to that of the bound state. Fig.~\ref{1halfpEall} presents analogous plots for various choices of the boundary condition (\(\zeta\)) in the \( 1/2^{+} \) channel. In each case, the GFMC evolution converges to a stable energy, which is then used in the direct method to determine the phase shift for the corresponding \( (E_{\text{c.m.}}, \zeta) \) pair.

%% file: Sections/Shifts.tex
Once the relative energy is known for a specified boundary condition in a given channel, Eq.~\eqref{tandSC} gives direct phase shifts. The relative energy also determines where to evaluate the surface amplitudes; for VMC, we compute Eqs.~(\ref{vmcassumefullA}) and (\ref{vmcassumefullB}), and for GFMC, we calculate Eqs.~(\ref{Aoftau}) and (\ref{Boftau}). The phase shifts in each channel for this system are given by $\tan\delta=B/A=\mathbf{\hat{K}}$ as detailed in Eq.~\eqref{kmat}. The two methods (direct and integral) should yield identical information and, therefore, we can use the direct method to validate the integral method for GFMC. The direct method in GFMC has already been successful for this system \cite{nollett2007quantum}. 
The main difference between the current and previous direct GFMC calculations is that, in the current method, we have explicitly removed scalar correlations from influencing the boundary condition. To benchmark our results, we also compare them with Faddeev-Yakubovsky (FY) calculations of $n+\alpha$ scattering using the AV18 potential~\cite{lazauskas2020description}. Additionally, we include phenomenological $R$-matrix calculations for each channel in our analysis~\cite{HaleHe5}.

\begin{figure}\centering
\includegraphics[width=12cm]{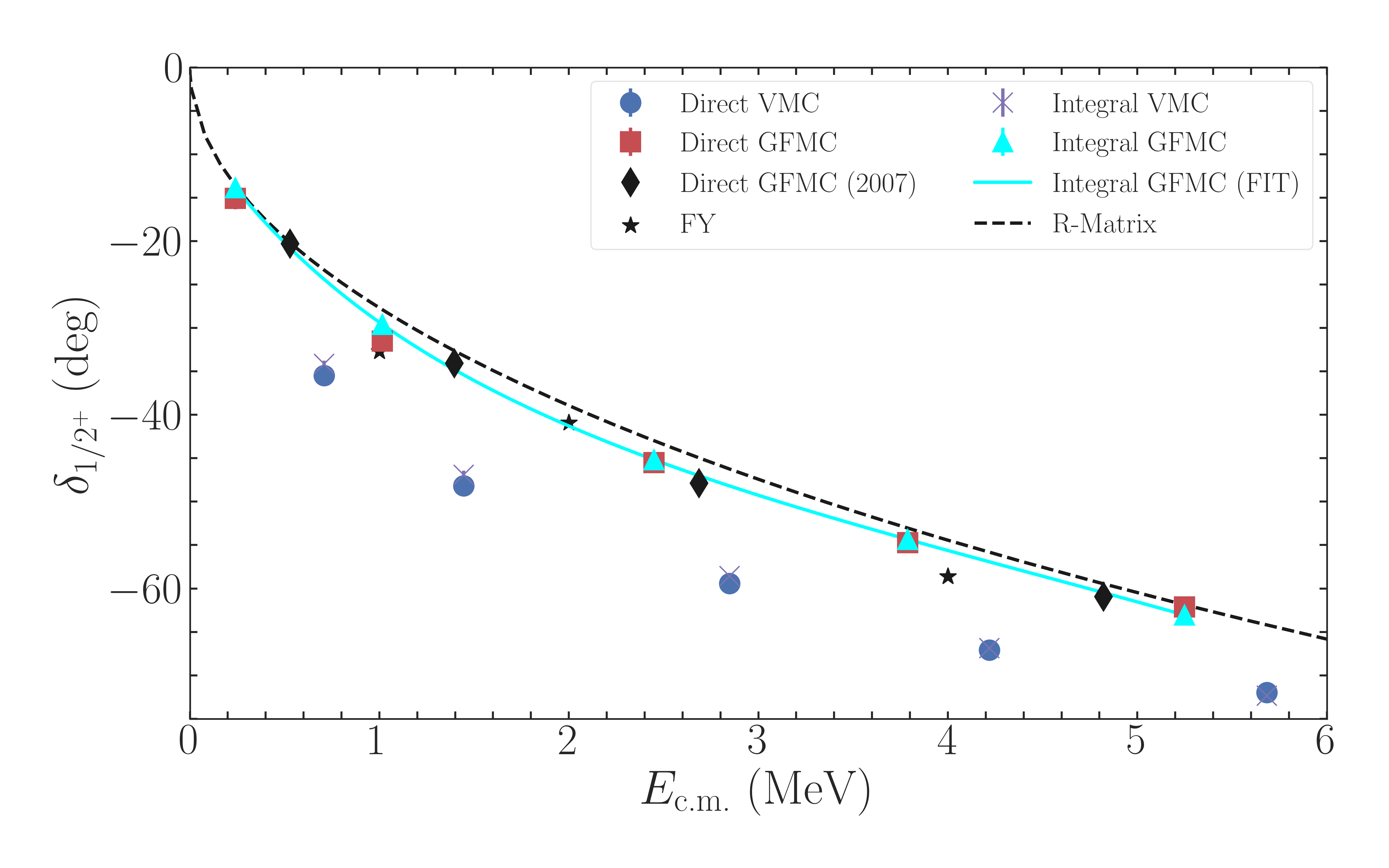}
\caption[]{$J^\pi=1/2^+$ phase shifts (in degrees) for $n+\text{}^4\text{He}$ calculated with multiple methods using the AV18 interaction. The phenomenological $R$-matrix phase shifts (dashed curve) are from Ref.~\cite{HaleHe5}. The Faddeev-Yakubovsky AV18 phase shifts (stars) are from Ref.~\cite{lazauskas2020description}. The previously computed direct GFMC results (diamonds) are from Ref.~\cite{nollett2007quantum}, which compares similarly to current direct GFMC results (squares) and the integral GFMC (triangles). We have also fit the integral GFMC results to a ratio of quadratic polynomials (solid curve). Lastly, we have the direct VMC (circles) and the integral VMC (x's) phase shifts.}\label{1halfpshifts}
\end{figure}

We begin by examining the \( s \)-wave (\( 1/2^{+} \)) phase shifts, shown in Fig.~\ref{1halfpshifts}. For each channel, we compare our results to a phenomenological \( R \)-matrix analysis of experimental data (dashed black curve), a Faddeev-Yakubovsky (FY) calculation (black stars), and a previously conducted direct GFMC calculation (black diamonds). We also plot a fit to a ratio of quadratic polynomials using the integral GFMC phase shifts (solid cure); this fit provides a straightforward way to visualize the agreement between the integral method and the other calculations. 

For the \( s \)-wave phase shifts, we find strong agreement between the current direct GFMC and the integral GFMC results (squares and triangles); we evaluated both methods at the same energy, and thus, we expect them to be within Monte Carlo sampling error of each other. Following the solid fitted curve, we find that previously computed direct GFMC  (diamonds) and the Faddeev-Yakubovsky (stars) phase shifts also firmly agree with the integral GFMC, a reassuring validation of the method. The close agreement between direct (circles) and integral (x’s) VMC results further confirms the correct implementation of corrections in the variational wave functions. Moreover, the necessity of imaginary time propagation becomes evident when computing the unbound system accurately—VMC phase shifts from either method do not align with our benchmarks, highlighting the limitations of the variational approach alone.

\begin{figure}\centering
\includegraphics[width=12cm]{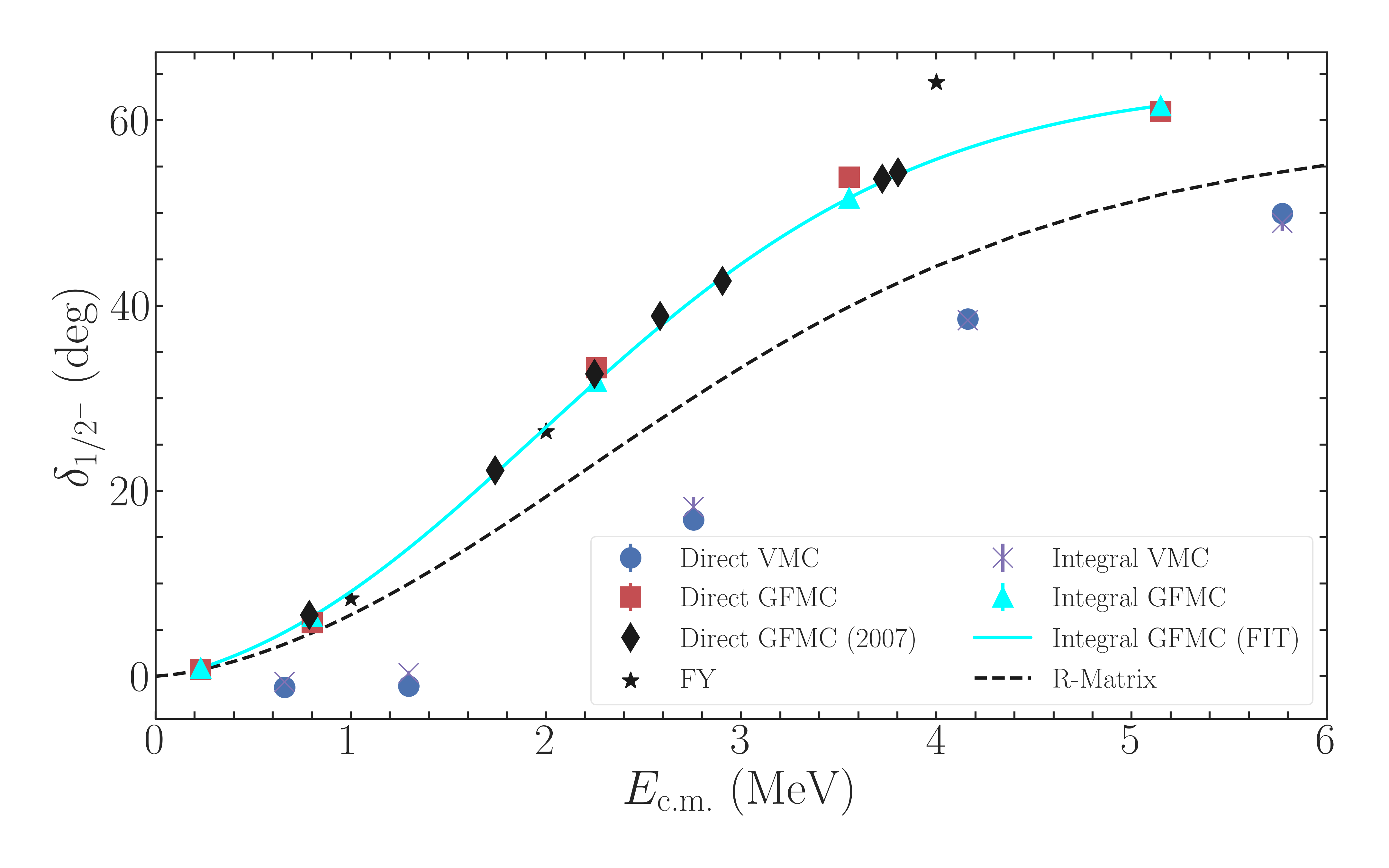}
\caption[]{$J^\pi=1/2^-$ phase shifts (in degrees) for $n+\text{}^4\text{He}$ calculated with multiple methods using the AV18 interaction, analogous to Fig.~\ref{1halfpshifts}.}\label{1halfmshifts}
\end{figure}

Moving on to the $p$-waves, in Fig.~\ref{1halfmshifts}, we examine the $1/2^{-}$ phase shift. At the computed energies, all GFMC results agree within the Monte Carlo sampling error, further validating the correct implementation of the integral method. However, the GFMC results (direct and integral) appear to conflict with the Faddeev-Yakubovsky result of Ref.~\cite{lazauskas2020description} at 4 MeV. This discrepancy seems to be caused by either a problem in the GFMC propagation (here and in Ref.~\cite{nollett2007quantum}) or an overestimated convergence of the FY phase shifts at this energy in all channels. Nonetheless, there is excellent agreement below 3 MeV between GFMC methods and FY calculations. Similarly to the \( 1/2^{+} \) case, there is a strong agreement between the direct (circles) and integral (x’s) VMC results. However, GFMC propagation remains essential for accurately benchmarking other theoretical calculations.

\begin{figure}\centering
\includegraphics[width=12cm]{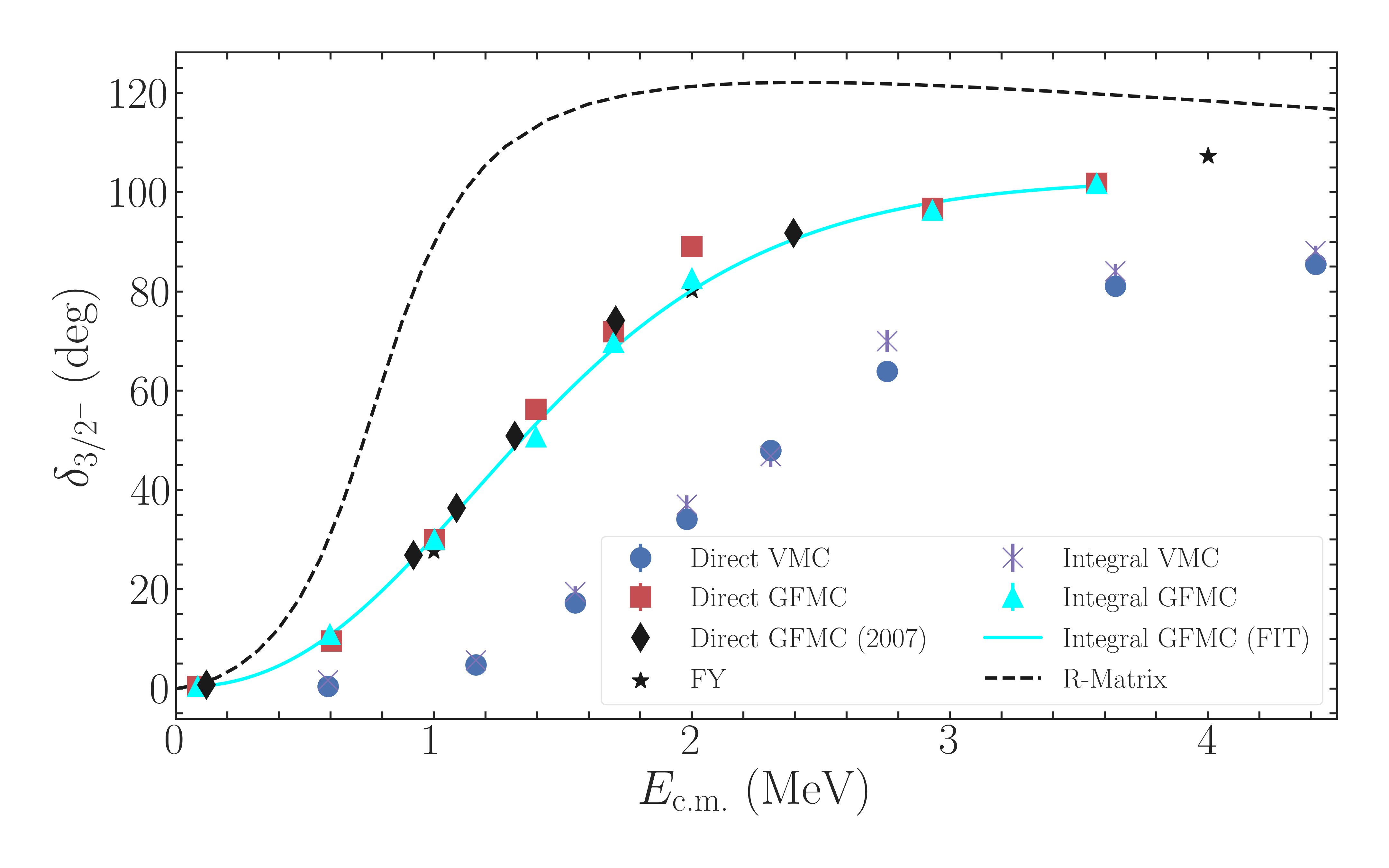}
\caption[]{$J^\pi=3/2^-$ phase shifts (in degrees) for $n+\text{}^4\text{He}$ calculated with multiple methods using the AV18 interaction, analogous to Fig.~\ref{1halfpshifts}.}\label{3halfmshifts}
\end{figure}

The other \( p \)-wave, the \( 3/2^{-} \) channel, exhibits a similar trend. As shown in Fig.~\ref{3halfmshifts}, the GFMC direct and integral methods, along with the FY calculations, agree within the sampling error. However, the direct GFMC phase shifts in this channel tend to fluctuate slightly between data points. This variation is likely due to the influence of the wave function tails in determining the system's energy; if the tails are not fully converged, small deviations in energy—on the order of a few degrees—can occur. Previous studies have also observed this effect, such as Refs.~\cite{nollett2007quantum, lynn2016Chiral}. In contrast, the integral method is expected to be less affected by this issue, as its dominant contributions come from the surface amplitudes at \( r_c < 9 \) fm, making it more resistant to uncertainties in the wave function tails. The most pronounced instance of this “jitter” in direct phase shifts appears in the \( 3/2^{-} \) channel at 2 MeV, where the integral method corrects for the fluctuation and aligns well with the FY calculation—an excellent validation of the integral approach.

%% file: Sections/Conclusion.tex
\setlength{\parskip}{0pt} 
\section{\label{section:conclusion}Conclusion}
In this study, we applied the integral method within the Green's Function Monte Carlo (GFMC) framework to calculate phase shifts for neutron-alpha ($n\text{-}\alpha$) scattering using the AV18 potential. By employing imaginary time propagation of wave functions, we obtained accurate $K$-matrix elements and phase shifts for the three dominant partial waves: $J^{\pi} = 1/2^{+}$ ($s$-wave), $J^{\pi} = 1/2^{-}$ ($p$-wave), and $J^{\pi} = 3/2^{-}$ ($p$-wave). Comparing our results with existing GFMC direct calculations~\cite{nollett2007quantum} and Faddeev-Yakubovsky solutions~\cite{lazauskas2020description} confirmed the statistical equivalence and robustness of the integral method.

A key advantage of the integral method over the direct method is its ability to reduce the influence of wave function tails, leading to more reliable scattering observables. Additionally, the integral method is readily adaptable to coupled-channel systems, making it a versatile tool for studying more complex nuclear reactions.

This work establishes the GFMC integral method as a systematic pathway for extending QMC calculations to single-nucleon scattering within the $p$-shell ($A \leq 12$). Furthermore, by combining the techniques developed here with the fixed interior wave approximation from Ref.~\cite{flores2023Var}, the method provides a systematic approach for extrapolating scattering lengths and threshold cross sections.

Building on these results, future applications of the integral method will explore coupled-channel scattering and the impact of three-body interactions on $n\text{-}\alpha$ scattering and other nuclear processes. This includes constraining low-energy constants within three-body potentials. Additionally, we plan to incorporate local chiral interactions~\cite{Piarulli:2016vel,Baroni:2018fdn} alongside the phenomenological potentials~\cite{Wiringa:1994wb,Pieper:2008rui} used in this work, providing a more consistent description of nuclear forces within the chiral effective field theory framework.

The development and application of the integral method in QMC offer a powerful tool for advancing {\it ab initio} calculations in nuclear reactions, paving the way for a unified framework to describe both bound and scattering states within the same theoretical approach.